\documentclass[12pt]{article}
\usepackage[latin1]{inputenc}
\usepackage[authoryear]{natbib}
\usepackage{graphicx}
\usepackage{amssymb, amsmath, amsthm}
\usepackage{times}
\usepackage{color}
\usepackage{bm}
\usepackage{multirow}
\usepackage{subfigure}
\usepackage{booktabs}

\newcommand{\vecbeta}{\boldsymbol{\beta}}
\newcommand{\vectheta}{\boldsymbol{\theta}}
\newcommand{\vecx}{\boldsymbol{x}}
\newcommand{\vecX}{\boldsymbol{X}}
\newcommand{\vecescore}{\boldsymbol{U}}
\newcommand{\vecW}{\boldsymbol{W}}
\newcommand{\vecy}{\boldsymbol{y}}

\newcommand{\Es}{{\rm E}}
\newcommand{\dd}{\mathrm{d}}
\newcommand{\diag}{{\rm diag}}
\newcommand{\SLR}{S_{\textrm{LR}}}
\newcommand{\SW}{S_{\textrm{W}}}
\newcommand{\SR}{S_{\textrm{R}}}
\newcommand{\ST}{S_{\textrm{T}}}
\newcommand{\LR}{{\textrm{LR}}}
\newcommand{\W}{{\textrm{W}}}
\newcommand{\R}{{\textrm{R}}}
\newcommand{\T}{{\textrm{T}}}

\title{Small-sample testing inference in symmetric and log-symmetric linear regression models}
\author{Francisco M.C.~Medeiros,\quad Silvia L.P.~Ferrari\footnote{Corresponding author: {\tt silviaferrari@usp.br}}\\
{\small {\em Department of Statistics, University of S\~ao Paulo, Brazil}}\\
}
\date{}

\begin{document}
\maketitle

\begin{abstract}
This paper deals with the issue of testing  hypothesis in symmetric and log-symmetric linear regression models in small and moderate-sized samples. 
We focus on four tests, namely the Wald, likelihood ratio, score, and gradient tests. These tests rely on asymptotic results and are unreliable
when the sample size is not large enough to guarantee a good agreement between the exact distribution of the test statistic and the corresponding 
chi-squared asymptotic distribution. Bartlett and Bartlett-type corrections typically attenuate the size distortion of the tests. These corrections 
are available in the literature for the likelihood ratio and score tests in symmetric linear regression models. Here, we derive a Bartlett-type 
correction for the gradient test. We show that the corrections are also valid for the log-symmetric linear regression models. We numerically 
compare the various tests, and bootstrapped tests, through simulations. Our results suggest that the corrected and bootstrapped tests exhibit 
type I probability error closer to the chosen nominal level with virtually no power loss. The analytically corrected tests, including the Bartlett-corrected
gradient test derived in this paper, perform as well as the bootstrapped tests with the advantage of not requiring computationally-intensive calculations. 
We present two real data applications to illustrate the usefulness of the modified tests.

\end{abstract}

%%%%%%%%%%%%%%%%%%%%%%%%%%%%%%%%%%%%%%%%%%%%%%%%%%%%%%%%%%%%%%%%%%%%%%%%%%%%%%%%%%%%%%%%%%%%%%%%%%%%%%%%%%%%%%%%%%%%%%%%%%%%%%%%%%%%%%%%%%%%
%                                                 Introduction
%%%%%%%%%%%%%%%%%%%%%%%%%%%%%%%%%%%%%%%%%%%%%%%%%%%%%%%%%%%%%%%%%%%%%%%%%%%%%%%%%%%%%%%%%%%%%%%%%%%%%%%%%%%%%%%%%%%%%%%%%%%%%%%%%%%%%%%%%%%%%

\section{Introduction}\label{sec:intro}

Normal linear regression models are widely employed in empirical research. There is a vast literature on extensions of these models to deal with non-normal errors.
In particular, much attention has been paid to symmetric linear and non-linear regression models; see \cite{Villegas-et-al-2013}, \cite{Lemonte2012}, \cite{PaulaCysneiros2009}, \cite{CysneirosPaulaGalea2007}, and \cite{Galea-et-al2005}. The idea is to replace the assumed normal distribution for the error by a wide class of symmetric distributions that encompasses distributions with heavier and lighter tails than the normal distribution. Some examples are the Student-t, type I logistic, type II logistic, contaminated normal, power exponential and slash distributions.

When dealing with positive, possibly skewed, data it is common practice to log-transform the observations and model the transformed data using a normal linear regression model. The corresponding model for the original data involves a multiplicative error term with a log-normal distribution. This limitation is relaxed in the class of the log-symmetric linear regression models  \citep{VanegasPaula2015a, VanegasPaula2015b}. It replaces the log-normal distribution by the class of log-symmetric distributions, which includes, for instance, the log-Student-t, type I log-logistic, type II log-logistic, log-contaminated-normal, log-power-exponential, and log-slash distributions.

Hypothesis testing in parametric models usually employ one of the three classic statistics, referred to as the ``Holy Trinity'' in the statistical literature, namely the Wald, likelihood ratio and Rao score statistics. Recently, the gradient statistic \citep{Terrell2002} has received attention. It is attractive because it is very simple to compute \citep[Section 1.8]{Rao2005}. It only involves the score vector and the maximum likelihood estimates (unrestricted and restricted to the null hypothesis) of the parameter vector. Unlike the Wald and score statistics, the gradient statistic does not require the information matrix, it does not involve matrix inversion, and it shares the same first order asymptotic properties with the other three statistics \citep{LemonteFerrari2012}. The four statistics differ in the second-order properties and may have considerably different behavior in finite samples.  

Let ${\cal H}_0$ be the null hypothesis to be tested against the alternative hypothesis ${\cal H}_1$. Under usual regularity conditions for likelihood inference,  the Wald, likelihood ratio, score and gradient statistics have an asymptotic $\chi^2_{q}$ distribution, where $q$ is the number of parameters fixed at ${\cal H}_0$. When the sample size is small or moderate, the $\chi^2_q$ approximation may be poor, and the true, unknown, type I probability error may be much larger than the nominal level of the test. Corrections to the test statistics have been proposed to attenuate this distortion. A Bartlett correction to the likelihood ratio statistic was first proposed by \cite{Bartlett1937} and later studied in full generality by \cite{Lawley1956}. \cite{CordeiroFerrari1991} derived a Bartlett-type correction to the score statistic. Recently, \cite{VargasFerrariLemonte2013} obtained a Bartlett-type correction to the gradient statistic. Typically, the order of approximation error of the null distributions of the statistics by the $\chi^2_q$ distribution is reduced from $O(n^{-1})$ to $O(n^{-2})$ when the corrections are applied. These corrections have been frequently employed in many parametric models; see, for instance, \cite{Chan2014}, \cite{Silva2014}, \cite{VargasFerrariLemonte2014}, \cite{BayerCribari2012}, \cite{LemonteFerrari2011}, \cite{Lemonte-et-al-2010}, \cite{Lagos-et-al-2010}, \cite{DamiaoCordeiro2009}, \cite{MeloFerrariCribari2009}, and \cite{BarrosoCordeiro2005}. 
Simulation experiments reported in these papers suggest that the size distortions of the tests may be substantially reduced in small and moderate-sized samples when the corrections are applied.   

Some works have focused on Bartlett and Bartlett-type corrections in symmetric regression models. \cite{Uribe2001} and \cite{Cordeiro2006} obtained a Bartlett correction to the likelihood ratio statistic for linear and nonlinear symmetric regression models, respectively. \cite{Uribe2007} and \cite{Cysneiros-et-al2010} obtained a Bartlett-type correction to the score statistic in linear and nonlinear symmetric regression models, respectively. In this paper, we derive a Bartlett-type correction to the gradient statistic in symmetric linear regression models. We show that the corrections derived for the symmetric linear regression models are also applicable to the log-symmetric linear regression models. Additionally, we present Monte Carlo experiments to study and compare the finite sample properties of the corrected and uncorrected tests and bootstrapped tests. 

This paper is organized as follows. In Section \ref{MRLS-sec:2} we define the symmetric linear regression models and discuss estimation and hypothesis testing issues. In Section \ref{MRLS-sec:3} we present a Bartlett correction to the likelihood ratio statistic and a Bartlett-type correction to the score statistic, and derive a Bartlett-type correction to the gradient statistic in the symmetric linear regression models. In Section \ref{MRLS-sec:4} we present the log-symmetric regression models and show that the 
results presented in the previous sections are valid in this class of models. In Section \ref{MRLS-sec:5} we present Monte Carlo simulation results. In Section \ref{MRLS-sec:6} we present two application in real data sets. Section \ref{MRLS-sec:7} closes the paper with final remarks.

%%%%%%%%%%%%%%%%%%%%%%%%%%%%%%%%%%%%%%%%%%%%%%%%%%%%%%%%%%%%%%%%%%%%%%%%%%%%%%%%%%%%%%%%%%%%%%%%%%%%%%%%%%%%%%%%%%%%%%%%%%%%%%%%%%%%%%%%%%%%
%                                    Symmetric linear regression models, estimation and testing
%%%%%%%%%%%%%%%%%%%%%%%%%%%%%%%%%%%%%%%%%%%%%%%%%%%%%%%%%%%%%%%%%%%%%%%%%%%%%%%%%%%%%%%%%%%%%%%%%%%%%%%%%%%%%%%%%%%%%%%%%%%%%%%%%%%%%%%%%%%%%

\section{Symmetric linear regression models, estimation and testing}\label{MRLS-sec:2}

Let $y$ be a continuous random variable having a symmetric distribution with location parameter $\mu \in \mathbb{R}$ and scale parameter $\phi>0$ and probability density function
\begin{equation}\label{mls}
\pi\left(y; \mu, \phi\right) =  \dfrac{1}{\phi}h\left(\left(\dfrac{y-\mu}{\phi}\right)^2\right),\hspace{1cm} y\in \mathbb{R},
\end{equation}
for some function $h:\mathbb{R}\rightarrow [0,\infty)$, such that $\int_{0}^{\infty}u^{-1/2}h(u)du=1$, called the density generating function. We write $y\sim S(\mu, \phi^2)$. Different choices for the function $h$ lead to different symmetric distributions; see Table \ref{tab1:1} for some examples. Two remarkable examples are the normal and the Student-t distributions. Note that $h$ may depend on extra parameters, for instance, the degrees of freedom parameter of the Student-t distribution. Whenever this is the case, such parameters are assumed to be fixed.

Some well-known properties of the normal distribution are valid for the symmetric distributions. For instance, if $y\sim S(\mu,\phi^2)$, then $a+by\sim S(a+b\mu, b^2\phi^2)$, with $a$, $b \in \mathbb{R}$, and $b\neq 0$. In particular, $z=(y-\mu)/\phi \sim S(0,1)$ and has probability density function $\pi(z; 0, 1)=h(z^2)$, $z\in\mathbb{R}$. Whenever they exist, $E(y)=\mu$ and $\mathrm{Var}(y)=\phi\xi$, where $\xi>0$ is a constant not depending on the parameters. The quantity $\xi$ for some distributions is presented in Table \ref{tab1:1}. Other results and properties of the symmetric distributions are given in \cite{BerkaneBentler1986}, \cite{Rao1990} and \cite{Fang-et-al1990}. 

Let $y_1, \ldots, y_n$ be independent random variables with $y_l\sim S(\mu_l, \phi^2)$, for $l=1, \dots n$. The symmetric linear regression models are defined as
\begin{equation}\label{e:mlsr}
y_l = \vecx_l^{\top}\vecbeta + \phi\,\epsilon_l,  \qquad  l=1,\ldots, n,
\end{equation}
where $\vecx_l=(x_{l1},\ldots, x_{lp})^\top$ is the vector of covariates associated to the $l$-th observation, $\vecbeta=(\beta_{1},\ldots, \beta_{p})^\top$ is a vector of unknown parameters and $\epsilon_1, \ldots, \epsilon_n$, are independent random errors with $\epsilon_l\sim S(0,1)$. We assume that $\vecX=(\vecx_1,\cdots,\vecx_n)$ is a full-rank $n\times p$ matrix, i.e. $\mathrm{rank}(\vecX)=p$. Additionally, we assume that usual regularity conditions for likelihood inference are valid \citep[Chap.9]{CoxHinkley1974}. The assumption that $\epsilon_l\sim S(0,1)$ relaxes the normality assumption for the errors allowing distributions with heavier tails (e.g. Student-t, type II logistic, power exponential with $0<k<1$) or lighter tails (e.g. type I logistic, power exponential with $-1<k<0$). The widely employed normal linear regression model is a special case of (\ref{e:mlsr}). 

The log-likelihood function for $\vectheta = (\vecbeta^\top,\phi)^\top$ is
\begin{equation}\label{e:vero}
l(\vectheta) = -n\log(\phi)+\displaystyle{\sum_{l=1}^{n}} g(z_l),
\end{equation}
with $g(z_l)=\log h(z_l^2)$ and $z_l=(y_l-\vecx_l^{\top}\vecbeta)/\phi$ being the standardized error associated to the $l$-th observation. The score vector for $\vecbeta$ and $\phi$ is given by $\vecescore(\vectheta) = (\vecescore_{\vecbeta}(\vectheta)^{\top}, U_{\phi}(\vectheta))^{\top}$ with
\[
 \vecescore_{\vecbeta}(\vectheta) = \phi^{-2}\vecX^{\top}\vecW(\vecy-\vecX\vecbeta), \ \ \
 U_{\phi}(\vectheta) = \phi^{-1}\left(\phi^{-2}(\vecy-\vecX\vecbeta)^\top\vecW(\vecy-\vecX\vecbeta) - n\right),
\]
\noindent where $\vecy=(y_1,\ldots,y_n)^{\top}$ and $\vecW=\diag\{w_1,\ldots,w_n\}$ with $w_l=-2\dd \log h(u)/\dd u|_{u=z_l^2}$.

The maximum likelihood estimates of $\vecbeta$ and $\phi$, obtained by  simultaneously solving the equations $\vecescore_{\vecbeta}(\vectheta)=\boldsymbol{0}$ and $U_{\phi}(\vectheta)=0$, are solutions of $\widehat{\vecbeta}=(\vecX^\top\widehat{\vecW}\vecX)^{-1}\vecX^\top\widehat{\vecW}\vecy$ and $\widehat{\phi}^2=n^{-1}\widehat{\boldsymbol{e}}^\top\widehat{\vecW}\widehat{\boldsymbol{e}}$ with $\widehat{\boldsymbol{e}}=\vecy-\vecX\widehat{\vecbeta}$ being the vector of the residuals. For each $l$, $w_l$ may be interpreted as the weight of the $l$-th observation in the estimation of the parameters. Table \ref{tab1:1} presents $w_l$ for some symmetric distributions. Note that the weights of all the observations are the same under the normality assumption for the errors. For the Cauchy, Student-t, type II logistic and power exponential (with $0<k<1$) distributions the weights $w_l$ are decreasing functions of $\left|z_l\right|$. Hence, the maximum likelihood estimators of $\vecbeta$ and $\phi$ are robust to the presence of outliers. For the type-I logistic and power exponential (with $-1<k<0$) distributions, the weights $w_l$ are increasing functions of $\left|z_l\right|$ because these  distributions have lighter tails than the normal distribution. 

\begin{table}[!htp]
{ \footnotesize
\begin{center}
\caption{Density generating function, $w$ and $\xi$ for some symmetric distributions.$^a$} \label{tab1:1}
\begin{tabular}{llllll}\hline
Distribution          & $h(u), \ u>0$                                                 	      & $w$                                   & $\xi$            
                      & $\delta_{20000}$      & $\delta_{20002}$ \\\hline
normal                & $\frac{1}{\sqrt{2\pi}}e^{-u/2}$              										      & $1$                                   & $1$
                      & $1$                   & $3$\\ &&&&&\\
Cauchy                & $\frac{1}{\pi}(1+u)^{-1}$                         										& $\frac{2}{1+z^2}$                     & does not exist
                      & $\frac{1}{2}$         & $\frac{3}{2}$ \\ &&&&&\\
Student-t             & $\frac{\nu^{\nu/2}}{B(1/2,\nu/2)}(\nu+u)^{-\frac{\nu+1}{2}},$         & $\frac{\nu+1}{\nu+z^2}$               & $\frac{\nu}{1+\nu}$, $\nu>2$
                      & $\frac{\nu+1}{\nu+3}$ & $\frac{3(\nu+1)}{\nu+3}$\\
											& $\nu>0$ &&&&\\ &&&&&\\                      
type I logistic       & $c\frac{e^{-u}}{(1+e^{-u})^2}, c\cong 1.4843$    										  & $2\tanh(z^2/2)$                       & $\cong0.79569$
                      & $\cong1.47724$        &$\cong4.01378$\\ &&&&&\\
type II logistic      & $\frac{e^{-\sqrt{u}}}{(1+e^{-\sqrt{u}})^2}$    											  & $\frac{e^{-|z|}-1}{|z|(1+e^{-|z|})}$  &$\frac{\pi^2}{3}$
                      & $\frac{1}{3}$         &$\cong2.42996$\\ &&&&&\\
%%exponencial dupla   &\exp{-\frac{1}{2}\sqrt(u)}/4$                                          & $1/2z$              &$16$\\
power exponential     & $\frac{1}{C(k)}\exp\{-\frac{1}{2}u^{1/(1+k)}\},$              				& $\frac{1}{(1+k)z^{2k/(1+k)}}$         &$2^{1+k}\frac{\Gamma(\frac{3}{2}(1+k))}{\Gamma(\frac{1+k}{2})}$
                      & $2^{1-k}\frac{\Gamma(\frac{3-k}{2})}{(1+k)^2\Gamma(\frac{1+k}{2})}$   & $\frac{3+k}{1+k}$\\
											& $-1<k\leq1$ &&&& \\ \hline
%                      & $C(k)=\Gamma(1+\frac{1+k}{2})2^{1+(1+k)/2}$                           & & & \\\hline
\multicolumn{6}{l}{{\small $^a$$B(\cdot, \cdot)$ and $\Gamma(\cdot)$ are the beta and gamma functions, respectively, and $C(k)=\Gamma(1+\frac{1+k}{2})2^{1+(1+k)/2}$}.}												
%\multicolumn{4}{l}{{\small $^a$$\Gamma(\cdot)$ is the gamma function.}}
\end{tabular}
\end{center}
}
\end{table}

Let $\delta_{abcde}=E(g^{(1)^a}g^{(2)^b}g^{(3)^c}g^{(4)^d}z^e)$, for $a,b,c,d,e \in \{0,1,2,3,4\}$, $g^{(r)}=\dd^r g(z)/\dd z^r$ and $z\sim S(0,1)$. \cite{Uribe2007} give the $\delta$'s for some symmetric distributions. The  $\delta$'s satisfy regularity relations such as $\delta_{01001}=\delta_{10000}=0$, $\delta_{20000}=-\delta_{01000}$, $\delta_{00010}=-\delta_{10100}$, $\delta_{40000}=-3\delta_{21000}$, $\delta_{01002}=2-\delta_{20002}$, $\delta_{11001}+\delta_{00101}+\delta_{01000}=0$, $2\delta_{00101}+\delta_{00012}+\delta_{10102}=0$, and $3\delta_{01002}+\delta_{11003}+\delta_{00103}=0$. The Fisher information matrix for $\vectheta=(\vecbeta^{\top}, \phi)^{\top}$ is block diagonal and is given by $\boldsymbol{K}=\diag\{\boldsymbol{K_\beta}, {K}_\phi\}$, with $\boldsymbol{K_\beta}=\phi^{-2}\delta_{20000}\vecX^\top\vecX$ and ${K}_\phi=n\phi^{-2}(\delta_{20002}-1)$. The quantities $\delta_{20000}$ and $\delta_{20002}$ are given in Table \ref{tab1:1} for some distributions. Hence, $\vecbeta$ and $\phi$ are globally orthogonal and their maximum likelihood estimators are asymptotically uncorrelated. 

The equations $\vecescore_{\vecbeta}(\vectheta)=\boldsymbol{0}$ and $U{\phi}(\vectheta)=0$ cannot be analytically solved, except for the normal model. The Fisher scoring iterative method for estimating $\vecbeta$ and $\phi$ may be implemented by iteratively solving 
\begin{align*}
\vecbeta^{(m+1)} &= \vecbeta^{(m)} + \frac{1}{\delta_{20000}}(\vecX^{\top}\vecX)^{-1}\vecX^{\top}\vecW^{(m)}(\vecy-\vecX\vecbeta^{(m)}), \\
\phi^{(m+1)} &= \phi^{(m)} + \frac{1}{\phi^{(m)}(\delta_{20002}-1)}\left(\frac{1}{n}(\vecy-\vecX\vecbeta^{(m)})\vecW^{(m)}(\vecy-\vecX\vecbeta^{(m)})-{\phi^{(m)}}^2\right),
\end{align*}   
for $m=0,1,\ldots$. The process may be initialized  with $\vecbeta^{(0)}=(\vecX^{\top}\vecX)^{-1}\vecX^{\top}\vecy$, the ordinary least squares estimate of $\vecbeta$, and $\phi^{(0)}=((\vecy-\vecX\vecbeta^{(0)})^{\top}(\vecy-\vecX\vecbeta^{(0)})/n)^{1/2}$. Some symmetric distributions can be obtained as a scale mixture of normal distributions, for example, the Student-t and power exponential ($0<k<1$) distributions \citep{AndrewsMallows1974, West1987}. Hence, the EM algorithm \citep{Dempster1977} may be used to find the maximum likelihood estimates of the parameters. Estimation in symmetric and log-symmetric models is implemented in the package {\tt ssym} \citep{VanegasPaulaR2015, VanegasPaulaRR2015} in {\tt R} \citep{R2015}.
 
Let $\mathcal{H}_{0}: \bm{\beta}_{1} = \bm{\beta}_{10}$ be the null hypothesis to be tested against $\mathcal{H}_{1}:\bm{\beta}_{1}\neq\bm{\beta}_{10}$, where the vector of unknown parameters $\bm{\beta}$ is partitioned as $\bm{\beta} = (\bm{\beta}_{1}^{\top}, \bm{\beta}_{2}^{\top})^{\top}$ with $\bm{\beta}_{1} = (\beta_{1}, \dots,\beta_{q})^{\top}$, $\bm{\beta}_{2} = (\beta_{q+1},\dots,\beta_{p})^{\top}$, and $\bm{\beta}_{10}$ is a $q$-vector of fixed constants.The partition of $\bm{\beta}$ induces the following partitions:  $\bm{U}_{\bm{\beta}} = (\bm{U}_{\bm{\beta}_1}^\top, \bm{U}_{\bm{\beta}_2}^\top)^\top$, with $\bm{U}_{\bm{\beta}_1}=\phi^{-2}\bm{X}_1^{\top}\vecW(\vecy-\vecX\vecbeta)$ and  $\bm{U}_{\bm{\beta}_2}=\phi^{-2}\bm{X}_2^{\top}\vecW(\vecy-\vecX\vecbeta)$, and
\[
\bm{K}_{\bm{\beta}} =
\begin{bmatrix}
\bm{K}_{\bm{\beta}11} & \bm{K}_{\bm{\beta}12} \\
\bm{K}_{\bm{\beta}21} & \bm{K}_{\bm{\beta}22}
\end{bmatrix} = \frac{\delta_{20000}}{\phi^2}
\begin{bmatrix}
\bm{X}_1^{\top}\bm{X}_1 & \bm{X}_1^{\top}\bm{X}_2 \\
\bm{X}_2^{\top}\bm{X}_1 & \bm{X}_2^{\top}\bm{X}_2
\end{bmatrix},
\]
with the matrix $\bm{X}$ partitioned as $\bm{X} = \bigl[\bm{X}_{1}\ \ \bm{X}_{2}\bigr]$, where $\bm{X}_{1}$ is an $n\times q$ matrix and $\bm{X}_{2}$ is an $n\times (p-q)$ matrix. To test $\mathcal{H}_{0}: \bm{\beta}_{1} = \bm{\beta}_{10}$ against $\mathcal{H}_{1}:\bm{\beta}_{1}\neq\bm{\beta}_{10}$ the Wald, likelihood ratio, score and gradient statistics may be employed. They are respectively given by 
\[
\SW = \delta_{20000}\widehat{\phi}^{-2}(\widehat{\bm{\beta}}_{1}- \bm{\beta}_{10})^{\top}
(\bm{R}^{\top}\bm{R})(\widehat{\bm{\beta}}_{1} - \bm{\beta}_{10}),
\]
\[
\SLR = 2\bigl\{\ell(\widehat{\bm{\beta}}_{1}, \widehat{\bm{\beta}}_{2},\widehat{\phi})
- \ell(\bm{\beta}_{10}, \widetilde{\bm{\beta}}_{2},\widetilde{\phi})\bigr\},
\]
\[
\SR = \frac{1}{\widetilde{\phi}^2\delta_{20000}}(\bm{y}-\vecX\widetilde{\vecbeta})^{\top}\widetilde{\bm{W}}\bm{X}_{1}
(\bm{R}^{\top}\bm{R})^{-1}\bm{X}_{1}^{\top}\widetilde{\bm{W}}(\bm{y}-\vecX\widetilde{\vecbeta}),
\]
\[
\ST = \widetilde{\phi}^{-2}(\bm{y}-\vecX\widetilde{\vecbeta})^{\top}\widetilde{\bm{W}}{\bm{X}}_{1}(\widehat{\bm{\beta}}_{1} - \bm{\beta}_{10}),
\]
where $(\widehat{\bm{\beta}}_{1}, \widehat{\bm{\beta}}_{2}, \widehat{\phi})$ and $(\bm{\beta}_{10}, \widetilde{\bm{\beta}}_{2}, \widetilde{\phi})$ are the unrestricted and restricted (to the null hypothesis) maximum likelihood estimator of  $(\bm{\beta}_{1}, \bm{\beta}_{2}, \phi)$, respectively, and $\bm{R} = \bm{X}_{1} - \bm{X}_{2}(\bm{X}_{2}^{\top}\bm{X}_{2})^{-1} \bm{X}_{2}^{\top}\bm{X}_{1}$. Tilde and hat are used to indicate evaluation at $(\widehat{\bm{\beta}}_{1}, \widehat{\bm{\beta}}_{2}, \widehat{\phi})$ and $(\bm{\beta}_{10}, \widetilde{\bm{\beta}}_{2}, \widetilde{\phi})$, respectively. Under the null hypothesis, the limiting distribution of the four statistics is $\chi_{q}^2$. Note that, unlike the Wald and score statistics, the gradient and likelihood ratio statistics do not involve matrix inversion.

%%%%%%%%%%%%%%%%%%%%%%%%%%%%%%%%%%%%%%%%%%%%%%%%%%%%%%%%%%%%%%%%%%%%%%%%%%%%%%%%%%%%%%%%%%%%%%%%%%%%%%%%%%%%%%%%%%%%%%%%%%%%%%%%%%%%%%%%%%%%
%                                 Improved tests in symmetric linear regression models
%%%%%%%%%%%%%%%%%%%%%%%%%%%%%%%%%%%%%%%%%%%%%%%%%%%%%%%%%%%%%%%%%%%%%%%%%%%%%%%%%%%%%%%%%%%%%%%%%%%%%%%%%%%%%%%%%%%%%%%%%%%%%%%%%%%%%%%%%%%%%

\section{Improved tests in symmetric linear regression models}\label{MRLS-sec:3}

Under usual regularity conditions, the statistics $\SW$, $\SLR$, $\SR$, and $\ST$ are asymptotically equivalent. In particular, they all have the same limiting $\chi^2_q$ distribution with approximation error of order $O(n^{-1})$ under $\mathcal{H}_{0}$. In small and moderate-sized samples, the use of the  $\chi^2$ approximation may cause considerable type I error probability distortion. Second-order asymptotic theory allows us to derive corrections to the test statistics that attenuate this problem. A Bartlett correction to the likelihood ratio statistic in symmetric linear regression models was obtained by \cite{Uribe2001}, and a Bartlett-type correction to the score statistic was derived by \cite{Uribe2007}. In this section, we derive a Bartlett-type correction to the gradient statistic. For this purpose, we make use of the general results in \cite{VargasFerrariLemonte2013}. Our results are new and represent a significant contribution to improvement of hypothesis testing in symmetric linear regression models.   

Define the following matrices: $\bm{Z}=\bm{X}(\bm{X}^{\top}\bm{X})^{-1}\bm{X}^{\top}$, $\bm{Z}_2=\bm{X}_2(\bm{X}_2^{\top}\bm{X}_2)^{-1}\bm{X}_2^{\top}$ (if $q<p$; $\bm{Z}_2=\bm{0}_{n\times n}$, if $q=p$), $\bm{Z}_d=\diag\{z_{11},\ldots,z_{nn}\}$, $\bm{Z}_{2d}=\diag\{z_{211},\ldots,z_{2nn}\}$. The matrices $\phi^{2}\delta_{20000}^{-1}\bm{Z}$ and $\phi^{2}\delta_{20000}^{-1}\bm{Z}_2$ are the asymptotic covariance matrices of $\bm{X}\widehat{\bm{\beta}}$ and $\bm{X}_2\widetilde{\bm{\beta}}_2$, respectively. Let $\rho_{ZZ}=n\mathrm{tr}(\bm{Z}_d \bm{Z}_d)$, $\rho_{Z_2Z_2}=n\mathrm{tr}(\bm{Z}_{2d} \bm{Z}_{2d})$, and $\rho_{ZZ_2}=n\mathrm{tr}(\bm{Z}_d \bm{Z}_{2d})$, where $\mathrm{tr}$ is the trace operator. From the general formulas in \cite{Lawley1956}, \cite{Uribe2001} obtained a Bartlett correction to the likelihood ratio statistic of $\mathcal{H}_{0}: \bm{\beta}_{1} = \bm{\beta}_{10}$ in symmetric linear regression models as
\begin{equation} \label{LRcor}
\SLR^{*}=\SLR(1-a_{\LR}),
\end{equation}
where $a_{\LR}$ is of order $O(n^{-1})$ and is given by $a_{\LR}=A_\LR+A_{\LR,\beta\phi}$, with
\begin{align*}
%A_{\LR}&=\frac{d_0}{q}\bm{1}_{n}^{\top}(\bm{Z}^2_{d}-\bm{Z}^2_{2d})\bm{1}_{n}^{\top},
A_{\LR}&=\frac{d_0}{nq}(\rho_{ZZ}-\rho_{Z_2Z_2}), 
\hspace{2cm} A_{\LR,\beta\phi}=\frac{d_1}{n}+\frac{d_2}{n}\frac{2p-q}{2},
\end{align*}
and
\begin{align*}
d_0=\frac{\delta_{00010}}{4\delta^2_{20000}}, \qquad d_1=-\frac{m_2m_3}{2m^2_1}-\frac{2m_3+m^2_3+m_4}{2m_1}, \qquad d_2=-\frac{m^2_3}{2m_1},
\end{align*}
where
\begin{align*}
m_1=\delta_{01002}-1, \hspace{0.5cm} m_2=4-\delta_{00103}-6\delta_{01002}, \hspace{0.5cm} m_3=\frac{\delta_{00101}+2\delta_{01000}}{\delta_{20000}}, \hspace{0.5cm} m_4=\frac{\delta_{00012}-6\delta_{11001}}{\delta_{20000}}.
\end{align*}
When $\phi$ is known, $A_{\LR,\beta\phi}=0$. Under the null hypothesis the corrected statistic $\SLR^{*}$ has an asymptotic $\chi_{q}^{2}$ distribution with approximation error of order $O(n^{-2})$. Hence, the correction factor reduces the approximation error from  $O(n^{-1})$ to $O(n^{-2})$. 
 
A Bartlett-type correction to the score test of $\mathcal{H}_{0}: \bm{\beta}_{1} = \bm{\beta}_{10}$ in symmetric linear regression models has been derived by  \cite{Uribe2007} from the general results of \cite{CordeiroFerrari1991}. The Bartlett-type corrected score statistic is given by
\begin{align} \label{escorecorr}
{S}_{\R}^{*}={S}_{\R}\bigl[1-\bigl({c}_{\R}+{b}_{\R}{S}_\R+{a}_{\R}{S}_{\R}^{2}\bigr)\bigr],
\end{align}
where ${a_\R}$, ${b_\R}$, and ${c_\R}$ are of order $O(n^{-1})$ and are given by ${a_\R}={A_{\R3}}/[12q(q+2)(q+4)]$, ${b_\R}=(A_{\R22} -2A_{\R3})/[12q(q+2)]$,
${c_\R}=(A_{\R11}-A_{\R22}+A_{\R3})/({12q})$, with
$A_{\R11}=A_{\R1}+A_{\R1,\beta\phi}$, $A_{\R22}=A_{\R2}+A_{\R2,\beta\phi}$,
\begin{align*}
A_{\R1}= \frac{12b_0}{n} (\rho_{ZZ_2}-\rho_{Z_2Z_2}), \qquad 
A_{\R2}=-\frac{9b_0}{n}(\rho_{ZZ}-2\rho_{ZZ_2}+\rho_{Z_2Z_2}), \qquad A_{\R3} =0,
\end{align*}
\[
A_{\R1,\beta\phi}=\frac{12b_1}{n}q(p-q)- \frac{6b_2}{n}q,\qquad
A_{\R2,\beta\phi}=-\frac{12b_3}{n}q(q+2),
\]
where 
\begin{align*}
b_0=\frac{\delta_{21000}}{\delta^2_{20000}}+1, \qquad b_1=\frac{\delta_{11001}(\delta_{11001}-\delta_{01000})}{\delta^2_{20000}(\delta_{20002}-1)}, \qquad b_3=\frac{\delta^2_{11001}}{\delta^2_{20000}(\delta_{20002}-1)}, 
\end{align*}
\begin{align*}
b_2=\frac{2\delta_{11001}(2\delta_{01002}+\delta_{00103})+(\delta_{20002}-1)(4\delta_{30001}+\delta_{40002}+\delta_{21002}-2\delta_{01000})}{\delta_{20000}(\delta_{20002}-1)^2}. 
\end{align*}
When $\phi$ is known, $A_{\R1,\beta\phi}=A_{\R2,\beta\phi}=0$. The correction factor $[1-({c}_{\R}+{b}_{\R}{S}_\R+{a}_{\R}{S}_{\R}^{2})]$ reduces to $[1-({c}_{\R}+{b}_{\R}{S}_{\R})]$ in \eqref{escorecorr} because $a_R=0$ in symmetric linear regression models. The asymptotic distribution of ${S}_{\R}^{*}$ is $\chi^2_q$ with the approximation error reduced from $O(n^{-1})$ to $O(n^{-2})$.

Recently, \cite{VargasFerrariLemonte2013} derived a Bartlett-type correction to the gradient statistic. The correction factor is a second-order polynomial in the gradient statistic. It diminishes the error of the $\chi^2$ approximation from $O(n^{-1})$ to $O(n^{-2})$. The results in \cite{VargasFerrariLemonte2013} are very general and need to be particularized for the parametric model and hypothesis of interest. The formulas involve moments of derivatives of the log-likelihood function up to the fourth order, and hence they may be hard or even impossible to obtain in many cases. In the following, we derive a closed-form expression for the Bartlett-type correction to the gradient statistic in symmetric linear regression models. 

From \cite{VargasFerrariLemonte2013} we obtained the Bartlett-type corrected gradient statistic for testing $\mathcal{H}_{0}: \bm{\beta}_{1} = \bm{\beta}_{10}$ in symmetric linear regression models as
\begin{align} \label{gradcorr}
{S}_{\T}^{*}={S}_{\T}\bigl[1-\bigl({c}_{\T}+{b}_{\T}{S}_\T+{a}_{\T}{S}_{\T}^{2}\bigr)\bigr],
\end{align}
where ${a_\T}$, ${b_\T}$, and ${c_\T}$ are of order $O(n^{-1})$ and are given by ${a_\T}={A_{\T3}}/[{12q(q+2)(q+4)}]$, ${b_\T}=(A_{\T22} -2A_{\T3})/[12q(q+2)]$, and
${c_\T}=(A_{\T11}-A_{\T22}+A_{\T3})/({12q})$, with
$A_{\T11}=A_{\T1}+A_{\T1,\beta\phi}$, and $A_{\T22}=A_{\T2}+A_{\T2,\beta\phi}$,
\begin{align*}
A_{\T1}= \frac{6c_0}{n} (\rho_{ZZ_2}-\rho_{Z_2Z_2}), \qquad
A_{\T2}=-\frac{3c_0}{n}(\rho_{ZZ}-2\rho_{ZZ_2}+\rho_{Z_2Z_2}), 
\qquad A_{\T3} =0,
\end{align*}
\[
A_{\T1,\beta\phi}=\frac{6c_1}{n}q(p-q) + \frac{6c_2}{n}q,\qquad
A_{\T2,\beta\phi}=-\frac{3c_1}{n}q(q+2),
\]
\[
c_0 = \frac{\delta_{00010}}{\delta^2_{20000}}, \qquad c_1=-\frac{m^2_3}{m_1}, \qquad c_2=-\frac{m_2m_3 + 2m_1m_3}{m^2_1}-\frac{m_4}{m_1}.
\]
When $\phi$ is known, $A_{\T1,\beta\phi}=A_{\T2,\beta\phi}=0$. The Bartlett-type correction factor $[1-({c}_{\T}+{b}_{\T}{S}_\T+{a}_{\T}{S}_{\T}^{2})]$ reduces to $[1-({c}_{\T}+{b}_{\T}{S}_{\T})]$ in \eqref{gradcorr} because ${a}_{\T} =0$. The derivation of these expressions is given in the Appendix.

The quantities $A_{\LR}$, $A_{\R1}$,  $A_{\R2}$, $A_{\T1}$, and $A_{\T2}$ do not depend on unknown parameters. They depend on the distribution assumed for the model error through the $\delta$'s and on the model matrix ${\bm X}$ through the diagonal elements of ${\bm Z}$ and ${\bm Z}_2$. The quantities $A_{\LR,\beta\phi}$, $A_{\R1,\beta\phi}$, $A_{\R2,\beta\phi}$, $A_{\T1,\beta\phi}$, and $A_{\T2,\beta\phi}$ represent the contributions generated by the fact that the scale parameter $\phi$ is unknown and estimated from the data.  Additionally, these quantities depend on the number of regression parameters ($p$), the number of parameters under test ($q$), and the distribution assumed for the data through the $\delta$'s. When computing $A_{\LR,\beta\phi}$, $A_{\R1,\beta\phi}$, $A_{\R2,\beta\phi}$, $A_{\T1,\beta\phi}$, and $A_{\T2,\beta\phi}$ the unknown $\phi$ may be replaced by its maximum likelihood estimate or any other consistent estimate. Although the $A$'s are all of order $O(n^{-1})$, they may be non-negligible in finite samples. 
It is noteworthy that the corrections are very simple and easily implemented in any software that performs simple matrix operations. 

It is necessary to obtain the $d$'s, $b$'s, and $c$'s for the chosen model to compute the corrected statistics. We now give these quantities for some symmetric distribution.

\paragraph{Normal.} $d_0=0, d_1=1, d_2=1, b_0=0, b_1=1, b_2=0, b_3=1/2, c_0=0, c_1=2, c_2=0$.
\vspace{0.3cm}

\paragraph{Student-t.}
\begin{align*}
\begin{split}
 d_0&=\dfrac{3(\nu+2)(\nu+3)^2}{2\nu(\nu+1)(\nu+5)(\nu+7)}, \quad d_1=\dfrac{(\nu+3)(\nu^3+11\nu^2+20\nu+4)}{\nu(\nu+7)(\nu+5)^2}, \quad d_2=\dfrac{(\nu+3)(\nu+2)^2}{\nu(\nu+5)^2},
\end{split}
\end{align*}
\begin{align*}
\begin{split}
 b_0&=\dfrac{6(\nu^2+4\nu-1)}{\nu(\nu+5)(\nu+7)},\quad b_1=\dfrac{(\nu-1)(\nu+2)(\nu+3)}{\nu(\nu+5)^2}, \quad b_2=-\dfrac{12(\nu^2+3\nu+2)(\nu+3)}{\nu(\nu+7)(\nu+5)^2},\\
 b_3&=\dfrac{(\nu-1)^2(\nu+3)}{2\nu(\nu+5)^2}, \quad c_0=\dfrac{6(\nu+2)(\nu+3)^2}{\nu(\nu+1)(\nu+5)(\nu+7)}, \quad c_1=\dfrac{2(\nu+2)^2(\nu+3)}{\nu(\nu+5)},\\
c_2&=-\dfrac{24(\nu+2)(\nu+3)}{\nu(\nu+7)(\nu+5)^2}.
\end{split}
\end{align*}

\paragraph{Type I logistic.} $d_0=-0.0767, d_1\approx1.4706, d_2\approx1.3626, b_0=-0.9035, b_1\approx1.7744, b_2\approx0.5690, b_3\approx1.1552, c_0=-0.3069, c_1\approx2.7253, c_2\approx0.2158$.
\vspace{0.3cm}

\paragraph{Type II logistic.} $d_0=3/20, d_1\approx0.7460, d_2\approx0.7867, b_0=2/5, b_1\approx0.5245, b_2\approx-0.5835, b_3\approx0.1748, c_0=3/5, c_1\approx1.5735, c_2\approx-0.0815$.
\vspace{0.3cm}

\paragraph{Power exponential ($-1<k<1/3$).}
\begin{align*}
\begin{split}
d_0&=\dfrac{k(1-k)\Gamma\left(\frac{1-3k}{2}\right)\Gamma\left(\frac{1+k}{2}\right)}{8\Gamma\left(\frac{3-k}{2}\right)^2}, \qquad d_1=d_2=\dfrac{1}{1+k},\qquad
b_0=1-\dfrac{(1-k)\Gamma\left(\frac{3-3k}{2}\right)\Gamma\left(\frac{1+k}{2}\right)}{2\Gamma\left(\frac{3-k}{2}\right)^2},\\
b_1&=\dfrac{1-k}{1+k}, \qquad b_2=\dfrac{2k(1-k)}{1+k}, \qquad b_3=\dfrac{(1-k)^2}{2(1+k)}, \qquad
c_0=\dfrac{k(1-k)\Gamma\left(\frac{1-3k}{2}\right)\Gamma\left(\frac{1+k}{2}\right)}{8\Gamma\left(\frac{3-k}{2}\right)^2},\\
c_1&=\frac{2}{1+k}, \qquad c_2=0.
\end{split}
\end{align*}

%%%%%%%%%%%%%%%%%%%%%%%%%%%%%%%%%%%%%%%%%%%%%%%%%%%%%%%%%%%%%%%%%%%%%%%%%%%%%%%%%%%%%%%%%%%%%%%%%%%%%%%%%%%%%%%%%%%%%%%%%%%%%%%%%%%%%%%%%%%%
%                                                 SEÇÃO MRLLS
%%%%%%%%%%%%%%%%%%%%%%%%%%%%%%%%%%%%%%%%%%%%%%%%%%%%%%%%%%%%%%%%%%%%%%%%%%%%%%%%%%%%%%%%%%%%%%%%%%%%%%%%%%%%%%%%%%%%%%%%%%%%%%%%%%%%%%%%%%%%%

\section{Log-symmetric linear regression models and improved tests}\label{MRLS-sec:4}

Let $t$ be a continuous random variable with density function
\begin{align}\label{dls}
\pi(t; \eta, \phi)=\frac{h(\tilde{t}^2)}{t\phi}, \qquad t>0,
\end{align}
where $\tilde{t} = \log\left[(t/\eta)^\frac{1}{\phi}\right]$, $\eta>0$ is the median of $t$, and $\phi>0$ is a shape (skewness or relative dispersion) parameter, for some function $h:\mathbb{R}\rightarrow [0,\infty)$, such that $\int_{0}^{\infty}u^{-1/2}h(u)du=1$. 
We write $t\sim LS(\eta, \phi^2)$. The distributions in (\ref{dls}) are called log-symmetric distributions because $\log(t)\sim S(\mu,\phi^2)$, with $\mu=\log(\eta)$. As before,  $h$ is the density generating function because different choices for $h$ lead to different distributions. Some special distributions in (\ref{dls}) are the log-normal, log-Student-t, type I log-logistic, type II log-logistic, and log-power-exponential distributions. This class of distributions is studied in \cite{VanegasPaula2015b} and provides a wide range of distributions to model continuous positive data.
A useful property of the log-symmetric distributions is that, if $\xi$ is a random variable with a standard log-symmetric distribution, i.e. $\xi\sim LS(1,1)$, and
\begin{align}\label{mrls}
t = \eta\,\xi^{\phi},
\end{align}
for some $\eta>0$ and $\phi>0$, we have $t\sim LS(\eta, \phi^2)$. Taking log on both sides of (\ref{mrls}) leads to the linearized equation $\log t = \log \eta + \phi\log(\xi)$,
and to the linear model defined below.

Let $t_1, \ldots, t_n$ be independent random variables with $t_l\sim LS(\eta_l, \phi^2)$, for $l=1,\ldots n$, $\vecx_l=(x_{l1},\ldots, x_{lp})^\top$ be a vector of covariates associated to the $l$-th observation and $\vecbeta=(\beta_{1},\ldots, \beta_{p})^\top$ be a vector of unknown parameters. The log-symmetric linear regression models are defined as 
\begin{align}\label{e:mrlls}
\log t_l = \log \eta_l + \phi\log(\xi_l), \qquad  l=1,\ldots n
\end{align}
where $\log \eta_l=\mu_l=\vecx_l^{\top}\vecbeta$, $\phi>0$ is unknown, and $\log(\xi_l)=\epsilon_l \sim S(0,1)$. Note that the median of $t_l$ is linearly related to the regression parameters through a log link function. A more general version of this model is defined and studied by \cite{VanegasPaula2015a}.

The regression model in (\ref{e:mrlls}) for $y_l=\log(t_l)$ is equivalent to the symmetric linear regression model (\ref{e:mlsr}). Hence, all the results in the previous sections are valid for model (\ref{e:mrlls}). In particular, the formulas for the Bartlett correction to the likelihood ratio statistic  given in (\ref{LRcor}), the Bartlett-type correction to the score statistic given in (\ref{escorecorr}), and the Bartlett-type correction to the gradient statistic given in (\ref{gradcorr}) are valid for the log-symmetric linear regression model (\ref{e:mrlls}). The results in the previous sections allow one to test hypotheses on the regression parameters using either the uncorrected statistics ($\SW$, $\SLR$, $\SR$, and $\ST$) or the corrected statistics ($\SLR^*$, $\SR^*$, and $\ST^*$).

%%%%%%%%%%%%%%%%%%%%%%%%%%%%%%%%%%%%%%%%%%%%%%%%%%%%%%%%%%%%%%%%%%%%%%%%%%%%%%%%%%%%%%%%%%%%%%%%%%%%%%%%%%%%%%%%%%%%%%%%%%%%%%%%%%%%%%%%%%%%
%                                                 Simulation results
%%%%%%%%%%%%%%%%%%%%%%%%%%%%%%%%%%%%%%%%%%%%%%%%%%%%%%%%%%%%%%%%%%%%%%%%%%%%%%%%%%%%%%%%%%%%%%%%%%%%%%%%%%%%%%%%%%%%%%%%%%%%%%%%%%%%%%%%%%%%%

\section{Simulation results}\label{MRLS-sec:5}

We now present a Monte Carlo simulation study to investigate and compare the performance of the Wald $(\SW)$, likelihood ratio $(\SLR)$, score $(\SR)$, and gradient $(\ST)$ tests and the corrected likelihood ratio $(\SLR^{*})$, score $(\SR^{*})$, and gradient $(\ST^{*})$ tests in small and moderate-sized samples in symmetric linear regression models. All the findings are valid for the log-symmetric linear regression models as well. Bootstrap versions of the tests are also included ($\SW^{b}$, $\SLR^{b}$, $\SR^{b}$, and $\ST^{b}$). 

We consider the model
\[
y_{l} = \beta_{0} + \beta_{1}x_{l1} + \cdots + \beta_{p-1}x_{l,p-1} + \phi\epsilon_l, \qquad l=1,\ldots,n,
\]
where $\epsilon_l$ are independent random errors. We consider the following distributions for the errors: standard normal, Student-t (with $\nu=4$ degrees of freedom), and type II logistic. The covariates $x_{l2}, \ldots, x_{lp}$ were taken as random draws from the ${\cal {U}}(0,1)$ distribution, all the regression parameters, except those fixed at the null hypothesis, equal 1, and the scale parameter is fixed at $\phi=3$. We considered different values for the number of regression parameters ($p$), the number of parameters under test ($q$), and the sample size ($n=20$, $25$, and $30$).

The number of Monte Carlo replicates is $15.000$ and the nominal levels are $\alpha = 10\%, 5\%$, and $1\%$. All the simulations were carried out in the matrix programming language {\tt Ox} (Doornik, 2013), that is freely available for academics purposes at {\tt http://www.doornik.com}. All the needed optimizations were performed using the quasi-Newton method BFGS using the library function {\sf MaxBFGS} with analytical derivatives. 

We evaluated through simulation the null rejection rates of $\mathcal{H}_0:\beta_1=\cdots=\beta_q=0$, i.e. the proportion of the time that each statistic ($\SW$, $\SLR$, $\SR$, $\ST$, $\SLR^{*}$, $\SR^{*}$, and $\ST^{*}$) is greater than the chosen $1-\alpha$ quantile of the $\chi^2_q$ reference distribution. The bootstrapped tests use the uncorrected statistics ($\SW$, $\SLR$, $\SR$, and $\ST$), and the critical points corresponding to the chosen nominal level are evaluated through parametric bootstrap with $600$ bootstrap replicates \citep{EfronTibshirani1993}. The results are presented in Tables \ref{tab1} and \ref{tab2} for the normal model, in Tables \ref{tab3} and \ref{tab4} for the Student-t model, and in Tables \ref{tab5} and \ref{tab6} for the type II logistic model.

Tables \ref{tab1}--\ref{tab6} suggest that the Wald test is markedly oversized, i.e. its type I error probability is much larger than the selected nominal level. For instance, for $p=6$, $n=20$, and $\alpha=5\%$ the null rejection rates for the normal model are $21.04\%$ ($q=4$) and $15.83\%$ ($q=2$); for the Student-t model we have $30.31\%$ ($q=4$) and $22.35\%$ ($q=2$), and for the type II logistic model we have $23.95\%$ ($q=4$), and $17.69\%$ ($q=2$); see Tables \ref{tab2}, \ref{tab4}, and \ref{tab6}, respectively. The likelihood ratio test presents null rejection rates well above the nominal levels, but it is less liberal than the Wald test. For instance, for the Student-t model with $p=6$, $q=2$, $n=20$, and $\alpha=5\%$ (Table \ref{tab4}), the null rejection rate of the likelihood ratio test is $14.09\%$ while the corresponding figure for the Wald test is $22.35\%$. The score and gradient tests seem to be more reliable than the likelihood ratio and Wald tests but still present some size distortion. Their null rejection rates in this setting are $7.77\%$ and $8.54\%$, respectively. For normal models, the null rejection rates of the score and gradient tests are exactly the same. We note that, for all the three models considered here, the null rejection rates of the score and gradient tests tend to increase as the number of  parameters under test decreases. Taken as a whole, the results in Tables \ref{tab1}--\ref{tab6} indicate that all the (uncorrected) tests may be substantially size distorted in small samples and, hence, corrections are needed.

The (analytically and bootstrap) corrected tests exhibit much smaller size distortion than the corresponding uncorrected tests. Hence, the analytical and bootstrap corrections are effective in bringing the type I error probability closer to the nominal significance level of the tests. Also, their null rejection rates are almost unaffected by the number of regression parameters or the number of parameters under test. As expected, corrections are not needed in large samples (simulation results not shown to save space) because the null rejection rates of all the tests approach the nominal level as the sample size increase.

\begin{table}[!htp]
\centering
{\footnotesize
\caption{Null rejection rates (\%) for $\mathcal{H}_0:\beta_{1}=\cdots=\beta_{q}=0$ with $p=4$ and $\phi=3$;
normal model.}    \label{tab1}
\begin{tabular}{lllrrrrrrrrrrr}  \hline
$q$&    $n$     & $\alpha(\%)$ & ${S}_\W$ & ${S}_{\LR}$ & ${S}_{\R}$ & ${S}_{\T}$ & ${S}^{*}_{\LR}$ & ${S}^{*}_{\R}$ & ${S}^{*}_{\T}$ &
${S}^{b}_\W$ & ${S}^{b}_{\LR}$ & ${S}^{b}_{\R}$ & ${S}^{b}_{\T}$\\\hline
\multirow{11}{*}{\centering 3}

 & \multirow{3}{*}{\centering 20}
  	        & 10    & 21.32 & 16.32 & 10.72  & 10.72 & 10.34 & 10.13 &  10.13 & 10.35 & 10.35 & 10.37 & 10.37\\
 &          & 5     & 14.63 & 9.51 &  4.33 & 4.33 & 5.10 & 4.63  & 4.63 & 5.12 & 5.15 & 5.15 & 5.15 \\
 &          & 1     & 6.12 & 2.62 &  0.37  & 0.37 & 1.09 & 0.73 & 0.73 & 1.13 & 1.17 & 1.19 & 1.19\\
 &          &       &       &       &       &       &       &       &  & & & &\\
 & \multirow{3}{*}{\centering 25}
		        & 10    & 18.94 & 14.83 &  10.49 & 10.49 & 10.22 & 10.09 &10.09  & 10.26 & 10.27 & 10.27 & 10.27\\
 &          & 5     & 12.16 & 8.32  & 4.65  & 4.65 & 5.22 & 4.93 & 4.93 & 5.33 & 5.35 & 5.36 & 5.36\\
 &          & 1     & 4.69 & 2.25 & 0.53  & 0.53 & 1.05 & 0.79 & 0.79 & 1.17 & 1.18 & 1.19 & 1.19\\
 &          &       &       &       &       &       &       &       &  & & & &\\
 & \multirow{3}{*}{\centering 30}
  	        & 10    & 16.97 & 13.60 & 10.23  & 10.23 & 9.97 & 9.85 & 9.85  & 9.99 &  10.00 & 10.01 & 10.01\\
 &          & 5     & 10.50 & 7.54 &  4.67 & 4.67 & 4.99 & 4.86 & 4.86 & 5.03 & 5.04 & 5.06 & 5.06\\
 &          & 1     & 3.73 & 1.77 & 0.49  & 0.49 & 0.96 & 0.78 & 0.78 & 1.03 & 1.03 & 1.04 & 1.04\\
 %\hline
 %\multirow{11}{*}{\centering 2}
     %
 %& \multirow{3}{*}{\centering 20}
			      %& 10    & 19.16 & 16.11 &  12.40 & 12.40 & 10.07 & 10.22 & 10.22 & 10.15 & 10.16 & 10.16 & 10.16\\
 %&          & 5     & 12.39 & 9.09 &  5.83 & 5.83 & 5.00 & 4.95 & 4.95 & 5.05 & 5.07 & 5.08 & 5.08\\
 %&          & 1     &  4.84 & 2.57 & 0.81  & 0.81 & 1.02 & 0.89 & 0.89 & 1.13 & 1.14 & 1.15 & 1.15\\
 %&          &       &       &       &       &       &       &       &  & & & & \\
 %& \multirow{3}{*}{\centering 25}
  	        %& 10    & 17.25 & 14.61 & 12.12  & 12.12 & 10.39 & 10.49 & 10.49 & 10.46 & 10.46 & 10.47 & 10.47 \\
 %&          & 5     & 10.86 & 8.41 &  5.81 & 5.81 & 5.07 & 5.06 & 5.06 & 5.14 & 5.15 & 5.16 & 5.16\\
 %&          & 1     &  3.89 & 2.25 &  0.79 & 0.79 & 1.00 & 0.91 & 0.91 & 1.03 &  1.05 &  1.05 &  1.05\\
 %&          &       &       &       &       &       &       &       &  & & & &\\
 %& \multirow{3}{*}{\centering 30}
  	        %& 10    & 15.65 & 13.67 & 11.37 & 11.37 & 9.78 & 9.82 & 9.82 & 9.82 & 9.83 & 9.83 & 9.83\\
 %&          & 5     & 9.17 & 7.34 & 5.40 & 5.40 & 4.85 & 4.83 & 4.83 & 4.93 & 4.93 & 4.94 & 4.94\\
 %&          & 1     & 3.07 & 1.79 & 0.79 & 0.79 & 0.95 & 0.88 & 0.88 & 1.01 & 1.02 & 1.03 & 1.03\\
 \hline
 \multirow{11}{*}{\centering 1}

 & \multirow{3}{*}{\centering 20}
			      & 10    & 16.23 & 14.69 & 13.24 & 13.24 & 9.98 & 10.22 & 10.22 & 10.05 & 10.05 & 10.07 & 10.07\\
 &          & 5     & 9.87 & 8.58 & 6.90 & 6.89 & 5.05 & 5.19 & 5.19 & 5.15 & 5.17 & 5.17 & 5.17\\
 &          & 1     & 3.73 & 2.43 & 1.29  & 1.29 & 1.04 & 1.02 & 1.02 & 1.11 & 1.12 & 1.14 & 1.14\\
 &          &       &       &       &       &       &       &       &  & & & & \\
 & \multirow{3}{*}{\centering 25}
  	        & 10    & 14.80 & 13.75 & 12.71 & 12.71 & 10.23 & 10.35 & 10.35 & 10.27 & 10.27 & 10.29 & 10.29\\
 &          & 5     & 8.82 & 7.74 & 6.60 & 6.61 & 5.23 & 5.31  & 5.31  & 5.29 & 5.31 & 5.31 & 5.31\\
 &          & 1     & 3.06 &  2.22 & 1.31 & 1.31 & 1.15 & 1.14 & 1.14 & 1.24 & 1.24 & 1.24 & 1.24\\
 &          &       &       &       &       &       &       &       &  & & & &\\
 & \multirow{3}{*}{\centering 30}
  	        & 10    & 13.79 & 12.87 & 12.02 & 12.02 & 10.08 & 10.13 & 10.13 & 10.11 & 10.11 & 10.11 & 10.11 \\
 &          & 5     & 8.19 & 7.27 & 6.38 & 6.38 & 5.16 & 5.23 & 5.23 & 5.24 & 5.24 & 5.24 & 5.24\\
 &          & 1     & 2.49 & 1.87 & 1.17 & 1.17 & 1.03 & 1.02 & 1.02 & 1.11 & 1.11 & 1.11 & 1.11 \\
 \hline
\end{tabular}      }
\end{table}

\begin{table}[!htp]
\centering
{\footnotesize
\caption{Null rejection rates (\%) for $\mathcal{H}_0:\beta_{1}=\cdots=\beta_{q}=0$ with $p=6$ and $\phi=3$;
normal model.}    \label{tab2}
\begin{tabular}{lllrrrrrrrrrrr}  \hline
$q$&    $n$     & $\alpha(\%)$ & ${S}_\W$ & ${S}_{\LR}$ & ${S}_{\R}$ & ${S}_{\T}$ & ${S}^{*}_{\LR}$ & ${S}^{*}_{\R}$ & ${S}^{*}_{\T}$
& ${S}^{b}_\W$ & ${S}^{b}_{\LR}$ & ${S}^{b}_{\R}$ & ${S}^{b}_{\T}$\\\hline
\multirow{11}{*}{\centering 4}

 & \multirow{3}{*}{\centering 20}
		       & 10    & 29.52 & 20.92 & 11.51 & 11.51 & 9.67 & 9.43 &  9.43 & 9.58 & 9.61 & 9.63 & 9.63\\
 &         & 5     & 21.04 & 12.89 & 4.56 & 4.56 & 4.80 & 4.33  & 4.33 & 4.79 & 4.79 & 4.79 & 4.79\\
 &         & 1     & 10.33 & 4.03 & 0.26 & 0.26 & 1.07 & 0.59 & 0.59 & 1.08 & 1.09 & 1.11 & 1.11\\
 &          &       &       &       &       &       &       &       &  & & & &\\
 & \multirow{3}{*}{\centering 25}
  	       & 10    & 24.43 & 17.93 & 10.97 & 10.97 & 9.71 & 9.56 & 9.56 & 9.67 & 9.67 & 9.67 & 9.67\\
 &         & 5     & 16.47  & 10.44 & 4.94 & 4.94 & 5.03 & 4.75 & 4.75 & 5.03 &  5.04 &  5.05 &  5.05\\
 &         & 1     & 7.44 & 3.17 & 0.38 & 0.38 & 0.94 & 0.64 & 0.64 & 1.01 & 1.03 & 1.06 & 1.06\\
 &          &       &       &       &       &       &       &       &  \\
 & \multirow{3}{*}{\centering 30}
  	       & 10    & 21.94 & 16.71 & 11.19 & 11.19 & 9.95 & 9.84 & 9.84 & 9.95 & 9.95 & 9.97 & 9.97\\
 &         & 5     & 14.29 & 9.43 & 4.97 & 4.97 & 5.00 & 4.78 & 4.78 & 5.02 & 5.02 & 5.03 & 5.03\\
 &         & 1     & 5.68 & 2.67 & 0.54 & 0.54 &  0.97 & 0.79 & 0.79 & 1.01 & 1.03 & 1.04 & 1.04\\

 \hline
 %\multirow{11}{*}{\centering 3}
     %
 %& \multirow{3}{*}{\centering 20}
			      %& 10    & 26.49 & 20.71 & 14.17 & 14.17 & 9.89 & 10.07 & 10.07 & 9.87 & 9.89 & 9.90 & 9.90\\
 %&          & 5     & 18.40 & 12.83 & 6.61 & 6.61 & 4.95 & 4.88 & 4.88 & 4.95 & 4.97 & 4.98 & 4.98\\
 %&          & 1     & 8.87 & 4.15 & 0.75 & 0.75 & 1.13 & 0.82 & 0.82 & 1.17 &  1.18 & 1.19 & 1.19\\
 %&          &       &       &       &       &       &       &       &  & & & & \\
 %& \multirow{3}{*}{\centering 25}
  	        %& 10    &  22.61 & 18.22 & 13.22 & 13.22 & 10.17 & 10.35 & 10.35 & 10.21 & 10.21 & 10.21 & 10.21\\
 %&          & 5     & 15.33 & 10.99 & 6.32 & 6.32 & 5.07 & 5.01 & 5.01 & 5.10 & 5.10 & 5.11 & 5.11\\
 %&          & 1     & 6.34 & 3.03 & 0.75 & 0.75 & 0.99 & 0.83 & 0.83 & 1.03 & 1.05 & 1.06 & 1.06\\
 %&          &       &       &       &       &       &       &       &  & & & &\\
 %& \multirow{3}{*}{\centering 30}
  	        %& 10    & 19.74 & 16.27 & 12.27 & 12.27 & 9.68 & 9.79 & 9.79 & 9.74 &  9.75 & 9.75 & 9.75\\
 %&          & 5     & 12.51 & 9.21 & 5.87 & 5.87 & 4.83 & 4.83 & 4.83 & 4.85 & 4.85 & 4.85 & 4.85\\
 %&          & 1     & 4.72 & 2.51 & 0.82 & 0.82 & 0.92 & 0.84 & 0.84 & 1.00 & 1.00 & 1.01 & 1.01 \\
  %\hline
 \multirow{11}{*}{\centering 2}
 
 & \multirow{3}{*}{\centering 20}
			      & 10    & 23.18 & 19.77 & 15.90 & 15.90 & 9.81 & 10.35 & 10.35 & 9.87 & 9.87 & 9.89 & 9.89\\
 &          & 5     & 15.83 & 12.26 & 8.25 & 8.25 & 4.89 & 5.11 & 5.11 & 4.95 & 4.95 & 4.95 & 4.95\\
 &          & 1     & 7.07 & 3.91 & 1.33 & 1.33 & 1.07 & 1.00 & 1.00 & 1.13 & 1.14 & 1.14 & 1.14\\
 &          &       &       &       &       &       &       &       &  & & & &\\
 & \multirow{3}{*}{\centering 25}
  	        & 10    & 19.88 & 17.21 & 14.38 & 14.38 & 10.14 & 10.39 & 10.39 & 10.18 & 10.19 & 10.19 & 10.19\\
 &          & 5     & 12.94 & 10.37  &  7.29 & 7.29 & 4.89 & 5.11 & 5.11 & 4.99 & 5.00 & 5.01 & 5.01\\
 &          & 1     & 4.99 & 2.90 & 1.24 & 1.24 & 1.00 & 0.97 & 0.97 & 1.07 & 1.09 & 1.10 & 1.10\\
 &          &       &       &       &       &       &       &       &  & & & &\\
 & \multirow{3}{*}{\centering 30}
  	        & 10    & 17.65 & 15.55 & 13.33 & 13.33 & 10.03 & 10.22 & 10.22 & 10.09 & 10.09 & 10.09 & 10.09\\
 &          & 5     & 11.19 &  9.07 & 6.85 & 6.85 & 5.07 & 5.15 & 5.15 & 5.11 & 5.11 & 5.13 & 5.13\\
 &          & 1     & 4.17 & 2.56 & 1.17 & 1.17 & 0.95 & 0.93 & 0.93 & 1.01 & 1.02 & 1.03 & 1.03\\
 \hline
\end{tabular}      }
\end{table}

%%%%%%%%%%%%%%%%%%%%%%%%%%%%%%%%%%%%%%%%%%%%%%%%%%%%%%%%%%%%%%%%%%%%%%%%%%%%%%%%%
%                     Student-t 
%%%%%%%%%%%%%%%%%%%%%%%%%%%%%%%%%%%%%%%%%%%%%%%%%%%%%%%%%%%%%%%%%%%%%%%%%%%%%%%%

\begin{table}[!htp]
\centering
{\footnotesize
\caption{Null rejection rates (\%) for $\mathcal{H}_0:\beta_{1}=\cdots=\beta_{q}=0$ with $p=4$ and $\phi=3$;
Student-t model.}    \label{tab3}
\begin{tabular}{lllrrrrrrrrrrr}  \hline
$q$&    $n$      & $\alpha(\%)$ & ${S}_\W$ & ${S}_{\LR}$ & ${S}_{\R}$ & ${S}_{\T}$ & ${S}^{*}_{\LR}$ & ${S}^{*}_{\R}$ & ${S}^{*}_{\T}$
& ${S}^{b}_\W$ & ${S}^{b}_{\LR}$ & ${S}^{b}_{\R}$ & ${S}^{b}_{\T}$\\\hline
\multirow{11}{*}{\centering 3}

 & \multirow{3}{*}{\centering 20}
		        & 10    & 26.57 & 17.53 & 11.55 & 10.67 & 10.33 & 10.42 & 10.06 & 10.45 & 10.45 & 10.64 & 10.54\\
 &          & 5     & 18.94 & 10.28 & 4.94 & 4.21 & 4.99 & 4.82 & 4.68 & 5.31 & 5.14 & 5.14 & 5.18\\
 &          & 1     & 9.24 & 2.81 & 0.52 & 0.33 & 0.99 & 0.77 & 0.74 & 1.05 & 1.07 & 1.12 & 1.13\\
 &          &       &       &       &       &       &       &       &  \\
 & \multirow{3}{*}{\centering 25}
  	        & 10    & 22.04 & 15.19 & 10.82 & 10.09 & 9.79 & 9.99 & 9.80 & 9.97 &  9.95 & 10.08 & 9.97\\
 &          & 5     & 14.73 & 8.58 & 5.01 & 4.64 & 5.20 & 4.95 & 5.03 & 5.25 & 5.33 & 5.19 & 5.29\\
 &          & 1     & 6.64 & 2.37 & 0.68 & 0.45 & 1.00 & 0.91 & 0.78 & 1.09 & 1.07 & 1.14 & 1.09\\
 &          &       &       &       &       &       &       &       &  \\
 & \multirow{3}{*}{\centering 30}
  	        & 10    & 19.78 & 14.23 & 10.58 & 10.04 & 9.73 & 9.86 & 9.65 & 9.74 & 9.89 & 9.97 & 9.88 \\
 &          & 5     & 12.69 & 7.79 & 4.91 & 4.54 & 4.82 & 4.85 & 4.87 & 5.04 & 4.88 & 5.01 & 5.09\\
 &          & 1     & 5.09 & 1.85 & 0.67 & 0.55 & 0.87 & 0.83 & 0.89 & 1.12 & 0.96 & 0.98 & 1.03\\
 %\hline
 %\multirow{11}{*}{\centering 2}
   %
 %& \multirow{3}{*}{\centering 20}
			      %& 10    & 23.27 & 17.15 & 12.73 & 12.89 & 10.22 & 10.21 & 10.23 & 10.47 & 10.47 & 10.14 & 10.37\\
 %&          & 5     & 15.81 & 10.11 & 6.55 & 6.09 & 5.21 & 5.29 & 5.07 & 5.18 &  5.37 & 5.34 & 5.28\\
 %&          & 1     & 7.22 & 2.84 & 0.99 & 0.74 & 0.92 & 0.91 & 0.82 & 1.04 & 1.07 & 1.12 & 1.11\\
 %&          &       &       &       &       &       &       &       &  \\
 %& \multirow{3}{*}{\centering 25}
  	        %& 10    & 19.63 & 14.98 & 11.80 & 11.83 & 9.80 & 10.09 & 9.95 & 9.93 & 9.97 & 10.08 & 10.06\\
 %&          & 5     & 12.75 & 8.51 & 5.68 & 5.59 & 5.00 & 4.86 & 4.91  & 5.19 & 5.13 & 4.99 &  5.09\\
 %&          & 1     & 5.34 & 2.47 & 0.99 & 0.75 & 1.00 & 1.01 & 0.95 & 1.11 & 1.14 & 1.16 & 1.17\\
 %&          &       &       &       &       &       &       &       &  \\
 %& \multirow{3}{*}{\centering 30}
  	        %& 10    & 18.06 & 14.26 & 11.55 & 11.73 & 10.03 & 10.12 & 10.11 & 10.01 & 10.14 & 10.12 & 10.17\\
 %&          & 5     & 11.37 & 7.95 & 5.69 & 5.51 & 4.82 & 4.93 & 4.87 & 5.06 & 4.99 & 5.01 & 5.01\\
 %&          & 1     & 4.09 & 2.09 & 1.00 & 0.87 & 1.01 & 0.98 & 0.94 & 1.22 & 1.15 & 1.09 & 1.09\\ 
  \hline
 \multirow{11}{*}{\centering 1}
 
 & \multirow{3}{*}{\centering 20}
			      & 10    & 18.51 & 15.79 & 13.35 & 13.79 & 10.20 & 10.48 & 10.37 & 10.37 & 10.35 & 10.36 & 10.39\\
 &          & 5     & 12.35 & 9.49 & 7.17 & 7.20 & 5.08 & 5.25 & 5.12 & 5.23 & 5.28 & 5.25 & 5.24\\
 &          & 1     & 4.97 & 2.71 & 1.23 & 1.19 & 0.86 & 0.90 & 0.89 & 1.04 & 1.07 & 1.09 & 1.03\\
 &          &       &       &       &       &       &       &       &  \\
 & \multirow{3}{*}{\centering 25}
  	        & 10    & 16.14 & 13.73 & 11.98 & 12.44 & 9.64 & 9.75 & 9.81 & 9.97 &  9.75 & 9.72 & 9.81\\
 &          & 5     & 10.05 & 7.80 & 6.03 & 6.39 & 4.79 & 4.86 & 4.83 & 4.81 & 4.88 & 4.87 & 4.91\\
 &          & 1     &  3.43 & 2.05 & 1.15 & 1.23 & 0.96 & 0.93 & 0.93 & 1.11 & 1.05 & 1.03 & 1.05\\
 &          &       &       &       &       &       &       &       &  \\
 & \multirow{3}{*}{\centering 30}
  	        & 10    & 14.93 & 13.01 & 11.65 &  11.99 & 9.61 & 9.75 & 9.76 & 9.77 &  9.77 & 9.74 & 9.78\\
 &          & 5     & 8.73 & 7.24 & 6.12 & 6.24 & 4.92 & 4.91 & 4.97 & 4.87 & 5.05 & 4.91 & 5.02\\
 &          & 1     & 2.83 & 1.84 & 1.31 & 1.31 & 1.05 & 1.05 & 1.09 & 1.22 & 1.11 & 1.11 & 1.16\\
 \hline
\end{tabular}      }
\end{table}

\begin{table}[!htp]
\centering
{\footnotesize
\caption{Null rejection rates (\%) for $\mathcal{H}_0:\beta_{1}=\cdots=\beta_{q}=0$ with $p=6$ and $\phi=3$;
Student-t model.}    \label{tab4}
\begin{tabular}{lllrrrrrrrrrrr}  \hline
$q$&    $n$      & $\alpha(\%)$ & ${S}_\W$ & ${S}_{\LR}$ & ${S}_{\R}$ & ${S}_{\T}$ & ${S}^{*}_{\LR}$ & ${S}^{*}_{\R}$ & ${S}^{*}_{\T}$
& ${S}^{b}_\W$ & ${S}^{b}_{\LR}$ & ${S}^{b}_{\R}$ & ${S}^{b}_{\T}$\\\hline
\multirow{11}{*}{\centering 4}

 & \multirow{3}{*}{\centering 20}
		        & 10    & 38.77 & 23.31 & 11.91 & 11.51 & 9.72 & 9.59 & 9.20 & 9.81 & 9.90 & 9.75 & 9.72\\
 &          & 5     & 30.31 & 14.41 &  4.98 & 4.53 & 4.76 & 4.37 & 4.39 & 4.87 & 4.99 & 4.81 & 4.94\\
 &          & 1     & 17.60 & 4.73 & 0.44 & 0.35 & 0.87 & 0.61 & 0.66 &  1.05 & 0.96 & 1.01 & 1.06\\
 &          &       &       &       &       &       &       &       &  \\
 & \multirow{3}{*}{\centering 25}
  	        & 10    & 31.99 & 20.18 & 12.16 & 11.71 & 10.03 & 10.28 & 9.89 & 10.01 & 10.21 & 10.40 & 10.27\\
 &          & 5     & 23.65 & 12.39 & 5.58 & 5.11 & 5.09 & 4.91 & 4.73 & 5.07 & 5.22 & 5.21 & 5.22\\
 &          & 1     & 12.33 & 3.85 & 0.64 & 0.47 & 0.95 & 0.79 & 0.77 & 1.13 & 1.05 & 1.11 & 1.13\\
 &          &       &       &       &       &       &       &       &  \\
 & \multirow{3}{*}{\centering 30}
  	        & 10    & 26.99 & 17.93 & 11.69 & 11.19 & 9.90  & 10.06 & 9.77 & 10.07 & 10.05 & 10.10 & 10.04\\
 &          & 5     & 18.99 & 10.31 & 5.37 & 4.90 & 4.99 & 4.89 & 4.67 & 5.07 & 5.17 & 5.02 & 4.97\\
 &          & 1     & 8.87 & 2.90 & 0.70 & 0.46 & 0.90 &  0.82 &  0.74 & 1.10 & 0.99 & 1.01 &  1.05\\
 %\hline
 %\multirow{11}{*}{\centering 3}
     %
 %& \multirow{3}{*}{\centering 20}
			      %& 10    & 35.30 & 23.46 & 14.09 & 14.59 & 9.61 & 10.23 & 9.76 & 9.90 & 9.95 & 9.99 & 9.77\\
 %&          & 5     & 26.91 & 14.39 & 6.62 & 6.41 &  4.63 & 4.73 & 4.62 & 4.81 & 4.85 & 4.87 & 4.83\\
 %&          & 1     & 14.91 & 4.66 & 0.89 & 0.66 & 0.78 & 0.78 &  0.73 & 0.93 & 0.85 & 1.01 & 1.02\\
 %&          &       &       &       &       &       &       &       &  \\
 %& \multirow{3}{*}{\centering 25}
  	        %& 10    & 29.04 & 20.01 & 13.41 & 13.60 & 9.90 & 10.37 & 10.15 & 10.03 & 10.17 & 10.33 & 10.31\\
 %&          & 5     & 20.97 & 12.23 & 6.86 & 6.57 & 4.80 & 5.13 & 4.97 & 5.09 & 5.04 & 5.25 & 5.21\\
 %&          & 1     & 10.31 & 3.77 & 1.03 & 0.83 & 0.90 & 0.90 & 0.83 & 1.06 & 1.07 & 1.17 & 1.06\\
 %&          &       &       &       &       &       &       &       &  \\
 %& \multirow{3}{*}{\centering 30}
  	        %& 10    & 24.57 & 17.77 & 12.95 & 13.13 & 9.97 & 10.37 & 9.99 & 10.03 & 10.17 & 10.30 & 10.05\\
 %&          & 5     & 16.66 & 10.29 & 6.39 & 6.29 & 4.87 & 4.95 & 4.87 & 5.01 & 5.03 & 4.99 & 5.05\\
 %&          & 1     & 7.50 & 2.77 & 1.06 & 0.85 & 0.97 & 0.97 & 0.85 & 1.11 & 1.07 & 1.11 & 1.06\\ 
  \hline
 \multirow{11}{*}{\centering 2}

 & \multirow{3}{*}{\centering 20}
			      & 10    & 30.09 & 22.26 & 15.61 & 17.10 & 9.86 & 10.29 & 10.48 & 10.28 & 10.23 & 9.97 & 10.23\\
 &          & 5     & 22.35 & 14.09 & 7.77 & 8.54 & 4.47 & 4.91 & 4.88  & 4.71 & 4.81 & 4.83 & 4.89\\
 &          & 1     & 11.71 & 4.37 & 1.32 & 1.20 & 0.78 & 0.85 & 0.73 & 0.87 & 0.91 & 0.98 & 0.97\\
 &          &       &       &       &       &       &       &       &  \\
 & \multirow{3}{*}{\centering 25}
  	        & 10    & 25.24 & 19.27 & 14.87 & 15.75 & 10.19 & 10.71 & 10.52 & 10.23 & 10.53 & 10.50 & 10.49\\
 &          & 5     & 17.69 & 11.99 & 7.78 & 8.11 & 5.01 & 5.33 & 5.19 & 5.11 & 5.21 & 5.33 & 5.29\\
 &          & 1     & 8.29 & 3.55 & 1.33 & 1.38 & 0.94 & 0.95 & 1.02 & 1.05 & 1.05& 1.12 & 1.13\\
 &          &       &       &       &       &       &       &       &  \\
 & \multirow{3}{*}{\centering 30}
  	        & 10    & 21.99 & 17.11 & 13.74 & 14.46 & 10.03 & 10.39 & 10.17 & 10.03 & 10.33 & 10.25 & 10.15\\
 &          & 5     & 14.55 & 9.97 & 6.81 & 7.29 & 5.05 & 5.06 & 5.23 & 5.23 &  5.23 & 5.05 & 5.26 \\
 &          & 1     & 6.12 & 2.91 & 1.31 & 1.35 & 0.98 & 0.94 & 0.93 & 1.15 & 1.09 & 1.07 & 1.15\\
 \hline
\end{tabular}      }
\end{table}

%%%%%%%%%%%%%%%%%%%%%%%%%%%%%%%%%%%%%%%%%%%%%%%%%%%%%%%%%%%%%%%%%%%%%%%%%%%%%%%%%
%                      Type II Logístic
%%%%%%%%%%%%%%%%%%%%%%%%%%%%%%%%%%%%%%%%%%%%%%%%%%%%%%%%%%%%%%%%%%%%%%%%%%%%%%%%

\begin{table}[!htp]
\centering
{\footnotesize
\caption{Null rejection rates (\%) for $\mathcal{H}_0:\beta_{1}=\cdots=\beta_{q}=0$ with $p=4$ and $\phi=3$;
type II logistic model.}    \label{tab5}
\begin{tabular}{lllrrrrrrrrrrr}  \hline
$q$&    $n$      & $\alpha(\%)$ & ${S}_\W$ & ${S}_{\LR}$ & ${S}_{\R}$ & ${S}_{\T}$ & ${S}^{*}_{\LR}$ & ${S}^{*}_{\R}$ & ${S}^{*}_{\T}$
& ${S}^{b}_\W$ & ${S}^{b}_{\LR}$ & ${S}^{b}_{\R}$ & ${S}^{b}_{\T}$\\\hline
\multirow{11}{*}{\centering 3}

 & \multirow{3}{*}{\centering 20}
		        & 10    & 22.69 & 16.28 & 10.99 & 10.42 &  9.68  & 9.90 & 9.83 & 10.25 & 10.15 & 10.11 & 10.15\\
 &          & 5     & 15.35 & 9.51 & 4.91 & 4.46 & 4.90 & 4.87 & 4.84 & 5.13 & 5.29 & 5.21 & 5.29\\
 &          & 1     & 6.95 & 2.77 & 0.39 & 0.35 &  0.96 & 0.77 & 0.71 & 1.14 & 1.11 & 1.11 & 1.10\\
 &          &       &       &       &       &       &       &       &  \\
 & \multirow{3}{*}{\centering 25}
  	        & 10    & 19.72 & 14.79 & 10.96 & 10.40 & 9.73 & 10.17 & 9.92 & 10.01 & 10.07 & 10.22 & 10.06\\
 &          & 5     & 12.71 & 8.23 & 5.09 & 4.74 & 5.07 & 5.06 & 5.02 & 5.17 & 5.26 & 5.22 & 5.25\\
 &          & 1     & 5.08 & 2.27 & 0.76 & 0.52 & 1.09 & 1.01 & 1.02 & 1.23 & 1.21 & 1.23 & 1.23\\
 &          &       &       &       &       &       &       &       &  \\
 & \multirow{3}{*}{\centering 30}
  	        & 10    & 17.53 & 13.73 & 10.63 & 10.09  & 9.51 & 9.96 & 9.73 & 9.79 & 9.81 & 10.01 & 9.91\\
 &          & 5     & 10.96 &  7.31 & 4.67 & 4.51 & 4.72 & 4.65 & 4.74 & 4.86 & 4.98 &  4.79 & 4.95\\
 &          & 1     & 3.91 & 1.71 & 0.54 & 0.49 & 0.77 & 0.74 & 0.68 & 0.93 & 0.88 & 0.92 & 0.91\\
 %\hline
 %\multirow{11}{*}{\centering 2}
   %
 %& \multirow{3}{*}{\centering 20}
			      %& 10    & 19.91 & 16.24 & 12.75 & 12.61 & 9.69 & 10.24 & 10.11 & 10.23 & 10.15 & 10.12 & 10.13\\
 %&          & 5     & 13.25 & 9.31 & 6.39 & 5.94 & 4.96 & 5.23 & 5.10 & 5.21 & 5.22 & 5.29 & 5.17\\
 %&          & 1     & 5.48 & 2.83 & 0.92 & 0.83 & 1.02 & 0.85 & 0.94 & 1.27 & 1.21 & 1.08 & 1.21\\
 %&          &       &       &       &       &       &       &       &  \\
 %& \multirow{3}{*}{\centering 25}
  	        %& 10    & 17.59 & 14.52 & 12.10 & 11.88 & 9.81 & 10.14 & 10.19 & 10.19 & 10.11 & 10.11 & 10.20\\
 %&          & 5     & 11.08 &  8.29 & 5.89 & 5.77 & 4.83 & 5.03 & 5.03  & 5.22 & 5.11 & 5.07 & 5.15\\
 %&          & 1     & 4.16 & 2.13 & 0.95 & 0.84 & 0.91 & 0.93 & 0.94 & 1.06 & 1.03 &1.07 & 1.09\\
 %&          &       &       &       &       &       &       &       &  \\
 %& \multirow{3}{*}{\centering 30}
  	        %& 10    & 15.95 & 13.25 & 11.50 & 11.35 & 9.49 & 9.92 & 9.85 & 9.87 & 9.84 &  9.88 & 9.87\\
 %&          & 5     & 9.67 & 7.26 & 5.57 & 5.41 & 4.54 & 4.91 & 4.73 & 4.86 & 4.81 & 4.97 & 4.81\\
 %&          & 1     & 3.20 & 1.79 & 0.91 & 0.79 & 0.87 & 0.89 & 0.85 & 0.99 & 0.97 &0.98 & 1.03\\ 
  \hline
 \multirow{11}{*}{\centering 1}
 
 & \multirow{3}{*}{\centering 20}
			      & 10    & 16.93 & 15.00 & 13.21 & 13.45  & 9.69 & 10.27 & 10.13 & 10.04 & 10.11 & 10.15 & 10.11\\
 &          & 5     & 10.47 & 8.71 & 6.97 & 6.93 & 4.58 & 4.95 & 4.92 & 4.95 & 4.93 & 4.93 & 4.95\\
 &          & 1     & 3.74 & 2.35 & 1.35 & 1.35 & 1.06 & 1.04 & 1.08 & 1.23 & 1.21 & 1.19 & 1.22\\
 &          &       &       &       &       &       &       &       &  \\
 & \multirow{3}{*}{\centering 25}
  	        & 10    & 15.09 & 13.75 & 12.50  & 12.76 & 9.62 & 10.16 & 9.93 & 9.89 & 9.93 & 10.12 & 9.91\\
 &          & 5     & 8.96 & 7.58 & 6.51 & 6.52 & 4.77 & 4.99 & 5.00 &  4.84 & 5.05 & 4.99 & 5.03\\
 &          & 1     & 2.83 & 1.88 & 1.22 & 1.19 & 0.94 & 0.92 & 0.99 & 1.08 & 1.09 & 0.97 & 1.06\\
 &          &       &       &       &       &       &       &       &  \\
 & \multirow{3}{*}{\centering 30}
  	        & 10    & 14.18  & 12.99 &  11.88 & 12.09 & 9.65 & 9.96 & 9.86 &  9.97 &  9.86 & 9.91 & 9.86\\
 &          & 5     & 8.21 & 6.94 & 6.05 & 6.08 & 4.65 & 4.91 & 4.87 & 4.82 & 4.89 &4.91 & 4.87 \\
 &          & 1     & 2.33 & 1.65 &  1.05 & 0.99 & 0.79 & 0.86 & 0.81 & 0.90 & 0.91 &  0.91 & 0.88\\
 \hline
\end{tabular}      }
\end{table}

\begin{table}[!htp]
\centering
{\footnotesize
\caption{Null rejection rates (\%) for $\mathcal{H}_0:\beta_{1}=\cdots=\beta_{q}=0$ with $p=6$ and $\phi=3$;
type II logistic model.}    \label{tab6}
\begin{tabular}{lllrrrrrrrrrrr}  \hline
$q$&    $n$      & $\alpha(\%)$ & ${S}_\W$ & ${S}_{\LR}$ & ${S}_{\R}$ & ${S}_{\T}$ & ${S}^{*}_{\LR}$ & ${S}^{*}_{\R}$ & ${S}^{*}_{\T}$
& ${S}^{b}_\W$ & ${S}^{b}_{\LR}$ & ${S}^{b}_{\R}$ & ${S}^{b}_{\T}$\\\hline
\multirow{11}{*}{\centering 4}

 & \multirow{3}{*}{\centering 20}
		        & 10    & 32.47 & 21.81 & 12.08 & 11.76 & 9.47 & 10.12 & 9.90 & 10.12 & 10.01 & 10.29 & 10.23\\
 &          & 5     & 23.95 & 13.57 & 5.03 & 4.82 & 4.79 & 4.61 & 4.63 & 5.24 & 5.26 & 5.11 & 5.30\\
 &          & 1     & 12.67 & 4.4 & 0.39 & 0.29 & 0.95 & 0.67 & 0.61 & 1.02 & 1.13 & 1.09 & 1.07\\
 &          &       &       &       &       &       &       &       &  \\
 & \multirow{3}{*}{\centering 25}
  	        & 10    & 26.35 & 18.37 & 11.85 & 11.79 & 9.88 & 9.93 & 10.22 & 10.25 & 10.37 & 9.95 & 10.38\\
 &          & 5     & 18.58 & 11.29 & 5.64 & 5.09 & 4.78 & 4.99 & 4.79 & 5.17 & 5.17  & 5.26 & 5.14\\
 &          & 1     & 8.87 & 3.32 & 0.61 & 0.46 & 0.94 & 0.84 & 0.78 & 1.18 & 1.09 & 1.15 & 1.15\\
 &          &       &       &       &       &       &       &       &  \\
 & \multirow{3}{*}{\centering 30}
  	        & 10    & 23.31 & 16.75 & 11.37 & 10.99 & 9.55 & 9.99 & 9.67 & 9.83 & 9.91 & 10.07 & 9.78\\
 &          & 5     & 15.63 & 9.65 & 4.89 & 4.80 & 4.57 & 4.50 & 4.66 & 5.07 & 4.93 & 4.71 & 4.94\\
 &          & 1     & 6.60 & 2.61 & 0.69 & 0.55 & 0.93 & 0.81 & 0.81 & 1.09 & 1.17 & 1.05 & 1.14\\
 %\hline
 %\multirow{11}{*}{\centering 3}
   %
 %& \multirow{3}{*}{\centering 20}
			      %& 10    & 29.27 & 21.79 & 14.33 & 14.27 & 9.40 & 10.15 & 10.15 & 10.04 & 10.08 & 9.94 & 10.08\\
 %&          & 5     & 21.21 & 13.19 & 6.75 & 6.49 & 4.59 & 4.94 & 4.89 & 5.05 & 4.99 & 5.08 & 5.11\\
 %&          & 1     & 10.37 & 4.35 & 0.81 & 0.78 & 0.93 & 0.81 & 0.86 & 1.11 & 1.13 & 1.11 & 1.14\\
 %&          &       &       &       &       &       &       &       &  \\
 %& \multirow{3}{*}{\centering 25}
  	        %& 10    & 24.07 & 18.21 & 13.29 & 13.25 & 9.53 & 10.11 & 10.09 & 10.16 & 9.94 & 9.90 & 10.05\\
 %&          & 5     & 16.39 & 10.81 & 6.69 & 6.57 & 4.97 & 5.17 & 5.19 & 5.23 & 5.31 & 5.21 & 5.36\\
 %&          & 1     & 7.45 & 3.29 & 0.98 & 0.82 & 0.92 & 0.91 & 0.83 & 1.00 & 1.11 & 1.11 & 1.15\\
 %&          &       &       &       &       &       &       &       &  \\
 %& \multirow{3}{*}{\centering 30}
  	        %& 10    & 21.38 & 16.36 & 12.19 & 12.17 & 9.05 & 9.79 & 9.55 & 9.62 & 9.55 &  & 9.55\\
 %&          & 5     & 13.65 & 9.06 & 6.07 & 5.87 & 4.61 & 4.87 & 4.70 & 4.89 & 4.79 & 4.94 & 4.83\\
 %&          & 1     & 5.34 & 2.71 & 1.08 & 0.97 & 1.05 & 0.99 & 1.00 & 1.12 & 1.17 & 1.15 & 1.21\\ 
  \hline
 \multirow{11}{*}{\centering 2}
 
 & \multirow{3}{*}{\centering 20}
			      & 10    & 25.27 & 20.30 & 15.61 & 16.25 & 9.34 & 10.52 & 10.55 & 10.07 & 10.19 & 10.09 & 10.23\\
 &          & 5     & 17.69 & 12.73 & 8.09 & 7.98 & 4.35 & 5.07 & 4.89 & 4.75 & 4.87 & 4.95 & 4.81\\
 &          & 1     & 7.95 & 3.95 & 1.39 & 1.26 & 0.87 & 0.99 & 0.91 & 1.10 & 1.10 & 1.13 & 1.01\\
 &          &       &       &       &       &       &       &       &  \\
 & \multirow{3}{*}{\centering 25}
  	        & 10    & 21.42 & 17.79 & 14.51 & 14.71 & 9.49 & 10.36 & 10.13 &  9.94 & 9.95 & 10.08 & 10.01\\
 &          & 5     & 14.11 & 10.41 & 7.55 & 7.45 & 4.73 & 5.08 & 5.11 & 5.07 & 5.09 & 4.94 & 5.09\\
 &          & 1     & 5.70 & 3.15 & 1.52 & 1.39 & 0.93 & 0.98 & 0.97 & 1.07 & 1.10 & 1.14 & 1.11\\
 &          &       &       &       &       &       &       &       &  \\
 & \multirow{3}{*}{\centering 30}
  	        & 10    & 19.37 & 16.19 & 13.40 & 13.55 & 9.53 & 10.13 & 9.95 & 9.81 & 9.93 &9.98  & 9.88\\
 &          & 5     & 12.25 & 9.17 & 6.76 & 6.76 & 4.54 & 4.89 & 4.78 & 4.86 & 4.83 & 4.89& 4.79\\
 &          & 1     & 4.37 & 2.57 &  1.27 & 1.21 & 0.87 & 0.99 & 0.96 & 1.04 & 0.99 & 1.08 & 1.03\\
 \hline
\end{tabular}      }
\end{table}

\begin{figure}[htbp!]
\centering
\subfigure[normal\label{fig:pri}]{\includegraphics[scale=0.38]{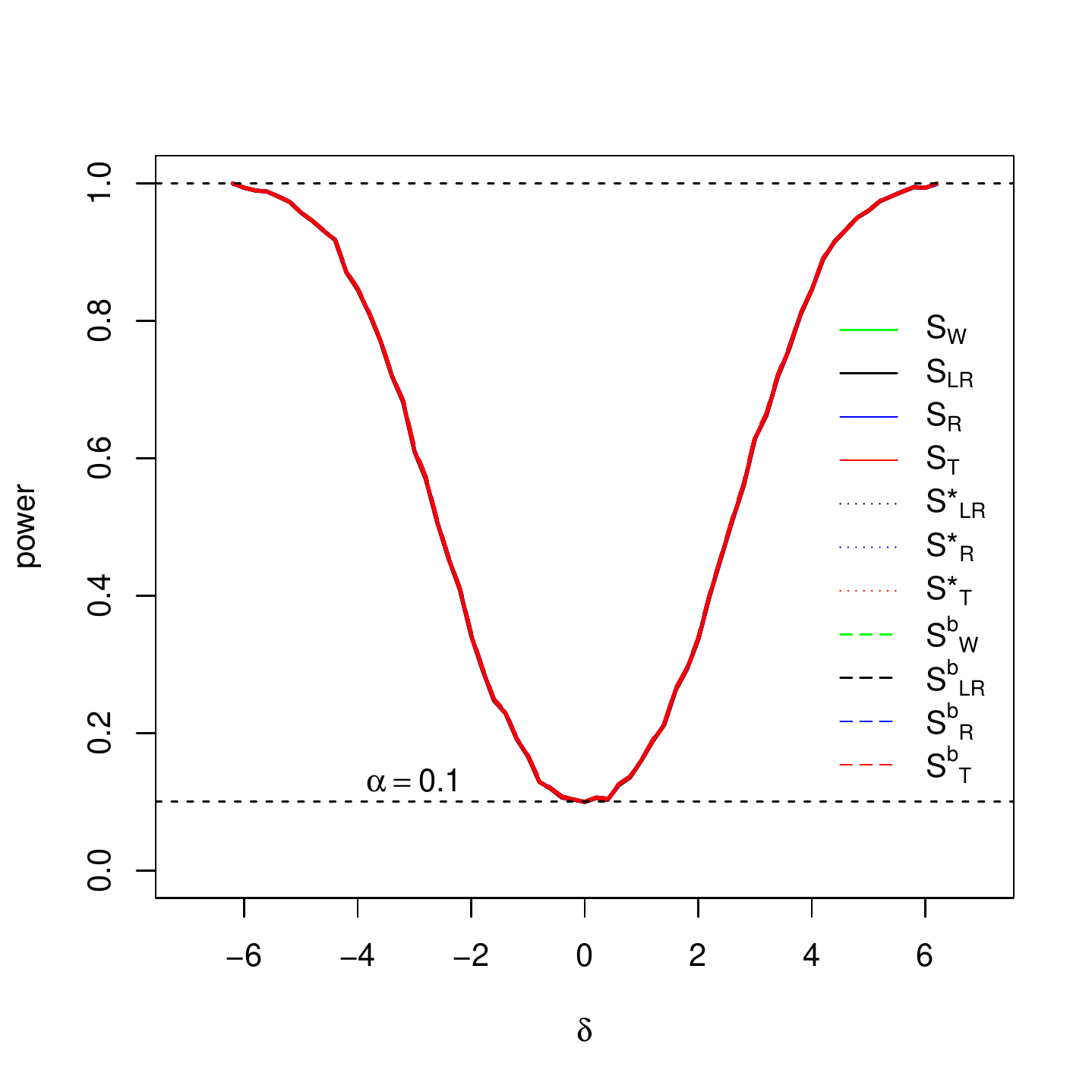}}\hfill
\subfigure[Student-t with $\nu=4$\label{fig:ter}]{\includegraphics[scale=0.38]{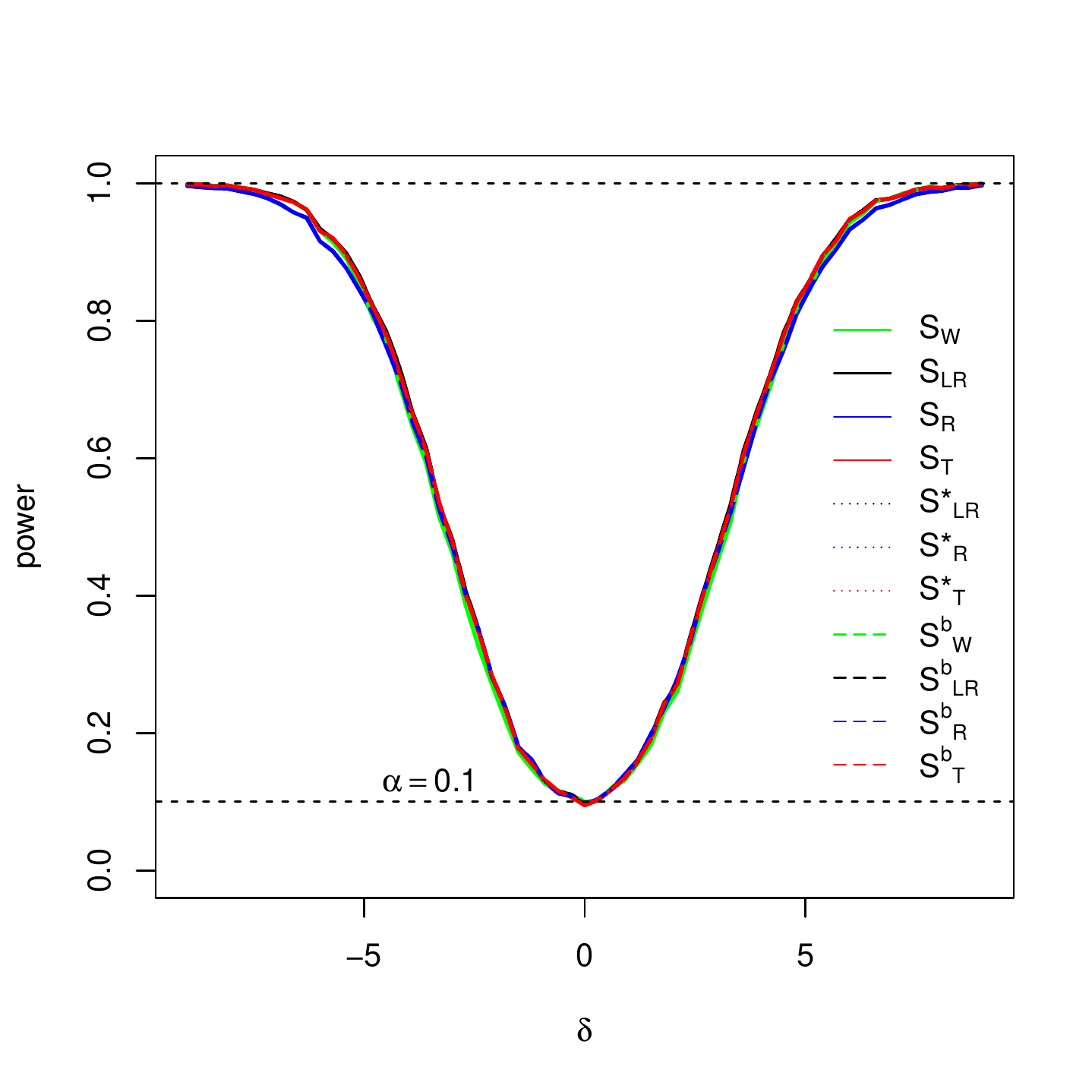}}\hfill
\subfigure[type II logistic\label{fig:qua}]{\includegraphics[scale=0.38]{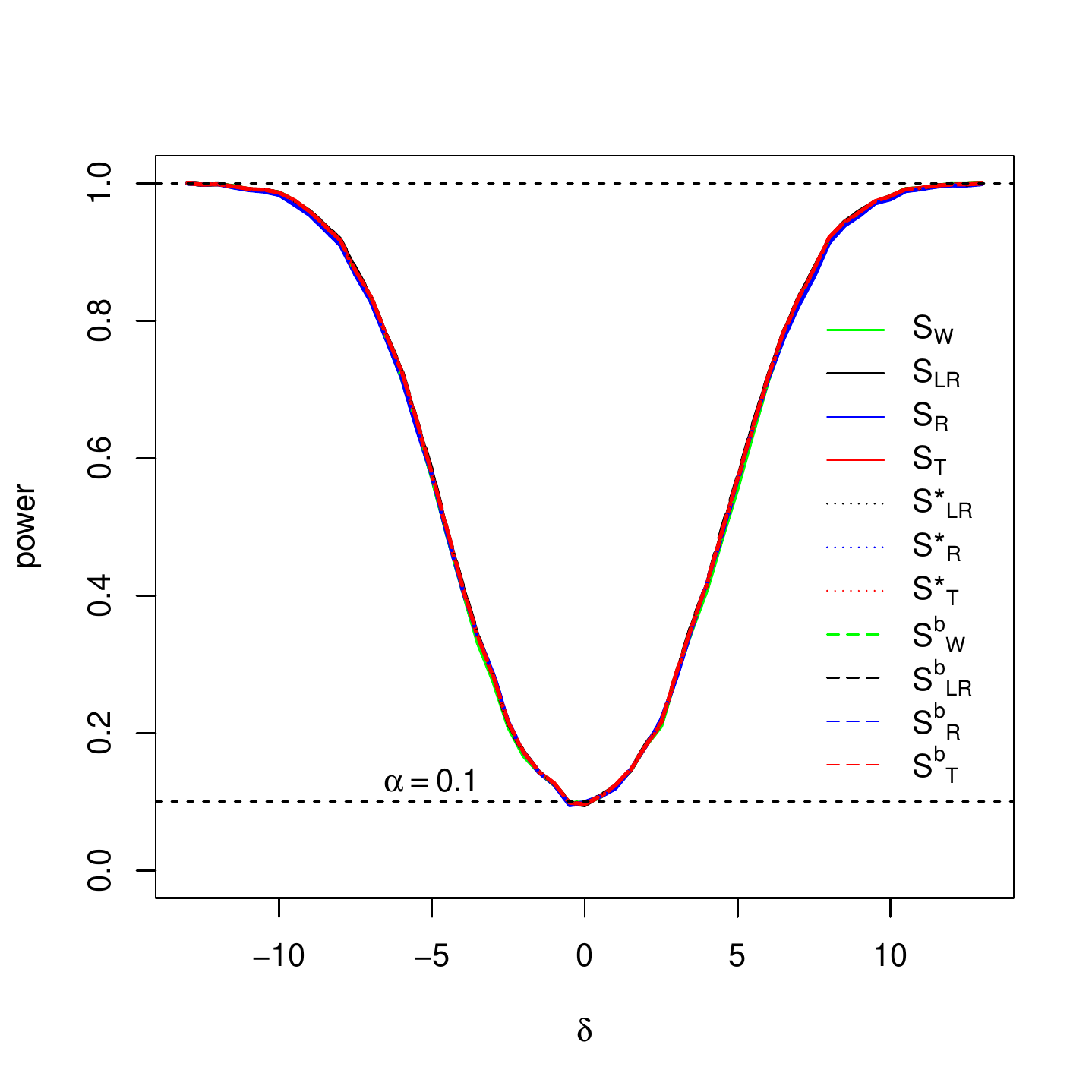}}\\
\caption{Power of the tests for $n=30$, $p=4$, $q=3$, $\phi=3$, and $\alpha=10\%$.}
\label{power}
\end{figure}

Now, the question is: ``Do the corrections induce power loss?'' From \cite{Lemonte2012}, we have that all the uncorrected and corrected tests have the same type I error probability and local power under Pitman alternatives up to an error of order $O(n^{-1})$ in symmetric and log-symmetric linear regression models. In the following, we numerically evaluate the power of all the tests in finite samples. Since the different tests have different sizes, we first generate $500.000$ Monte Carlo samples to estimate the critical value of each test that guarantees the correct significance level. This strategy can be applied to all the tests that use a critical value, but not the bootstrapped tests. We considered the null hypothesis $\mathcal{H}_0:\beta_1=\beta_2=\ldots=\beta_q=0$, and computed the rejection rates under the alternative hypothesis $\mathcal{H}_1:\beta_1=\beta_2=\ldots=\beta_q=\delta$. Figure \ref{power} plots the power of the tests  as a function of $\delta$, with $n=30$, $p=4$, $q=3$, $\phi=3$, and $\alpha=10\%$ for the normal, Student-t, and type II logistic models. Note that the bootstrapped tests have significance levels close to $\alpha$ in these situations; see Tables \ref{tab1}, \ref{tab3}, and \ref{tab5}. The curves are almost indistinguishable, and reveal that the all the tests (corrected and uncorrected) have similar powers. As expected, the power tends to 1 as $|\delta|$ grows. Power simulations for different values of $n$, $p$, $q$, $\phi$, and $\alpha$ (not shown) exhibited a similar pattern.

Overall, the Monte Carlo simulation results reveal that, in small and moderate-sized samples, the Wald, likelihood ratio, score and gradient tests tend to be liberal, i.e. they wrongly reject the null hypothesis more frequently than allowed by the chosen nominal significance level. Among all the tests we considered, the Wald test is clearly the most liberal. The Bartlett and Bartlett-type corrections are effective in correcting the size distortions of the tests with virtually no power loss. The bootstrapped tests perform similarly  to the analytically corrected tests at the cost of requiring computationally-intensive calculations. There is no Bartlett-type correction available to the Wald statistic and, hence, the bootstrap method is a convenient tool to correct its liberal behavior. We conclude that the modified (analytically corrected or bootstrapped) tests are to be preferred for testing hypotheses in symmetric and log-symmetric linear regression models when the sample is small or of moderate size.

%%%%%%%%%%%%%%%%%%%%%%%%%%%%%%%%%%%%%%%%%%%%%%%%%%%%%%%%%%%%%%%%%%%%%%%%%%%%%%%%%%%%%%%%%%%%%%%%%%%%%%%%%%%%%%%%%%%%%%%%%%%%%%%%%%%%%%%%%%%%
%                                              Applications
%%%%%%%%%%%%%%%%%%%%%%%%%%%%%%%%%%%%%%%%%%%%%%%%%%%%%%%%%%%%%%%%%%%%%%%%%%%%%%%%%%%%%%%%%%%%%%%%%%%%%%%%%%%%%%%%%%%%%%%%%%%%%%%%%%%%%%%%%%%%%

\section{Applications}\label{MRLS-sec:6}

\begin{figure}[!ht]
\centering
\subfigure[log-normal\label{fig:pri_queijo}]{\includegraphics[scale=0.35]{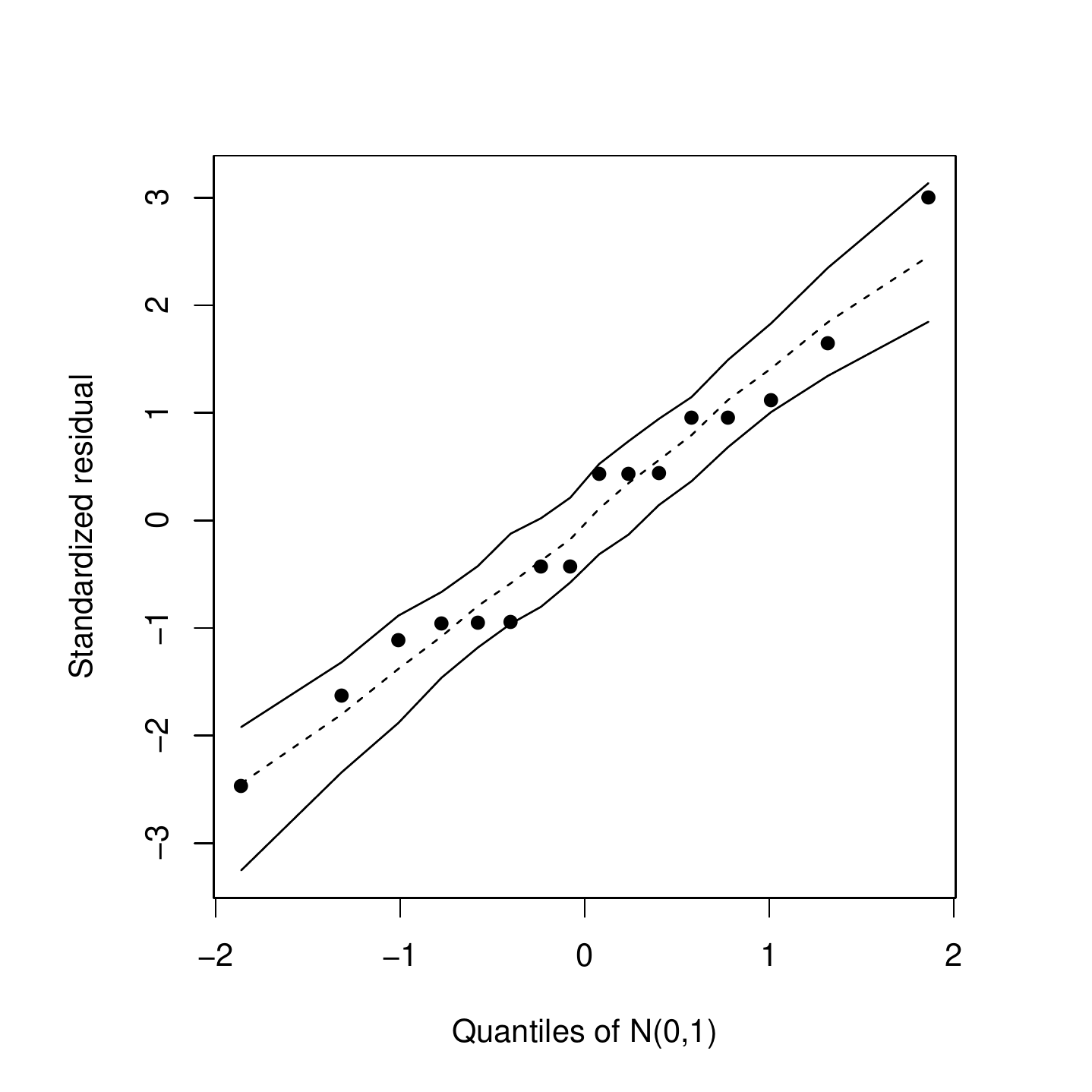}}\hspace{0.6cm}
\subfigure[log-Student-t\label{fig:seg_queijo}]{\includegraphics[scale=0.35]{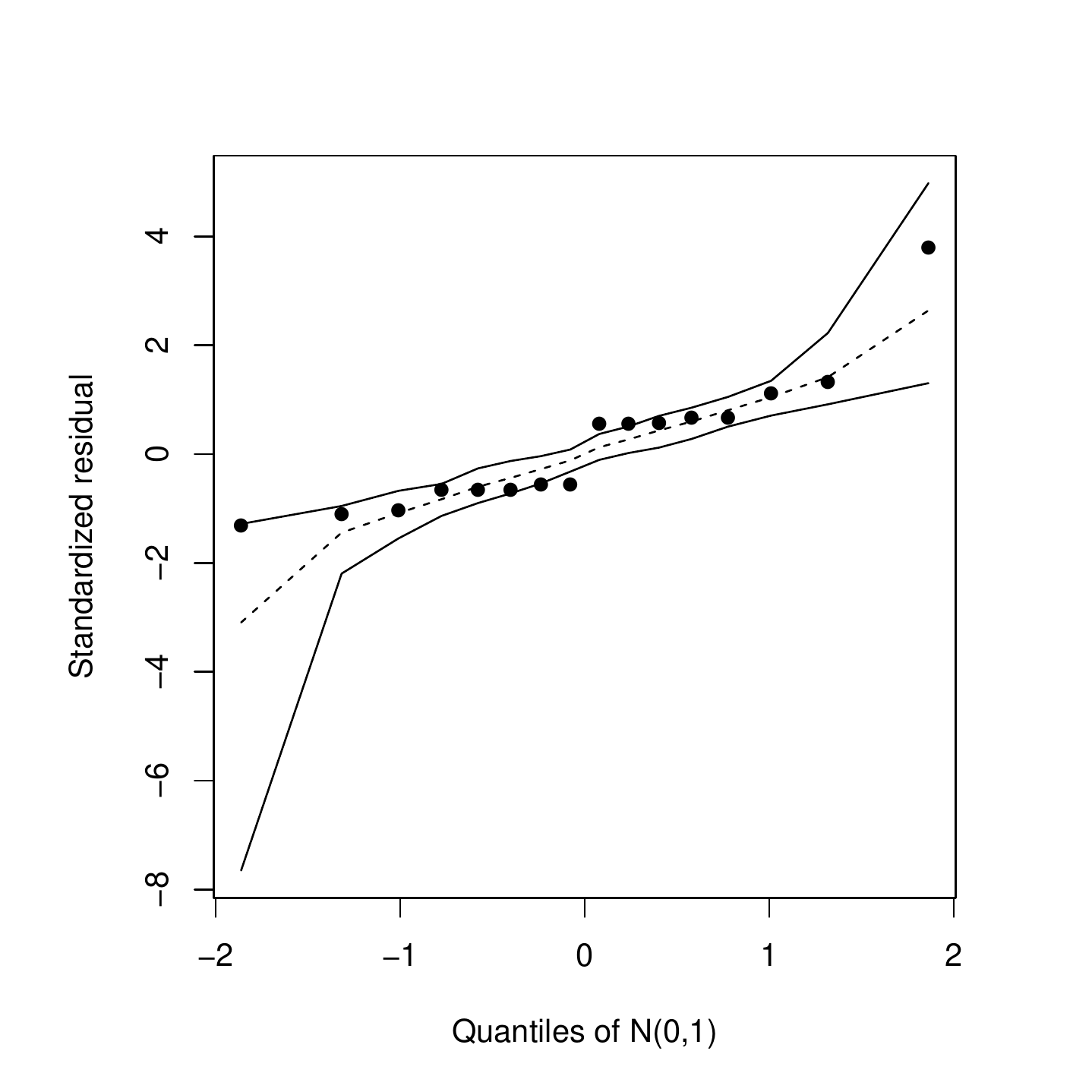}}
\subfigure[type II log-logistic\label{fig:ter_queijo}]{\includegraphics[scale=0.35]{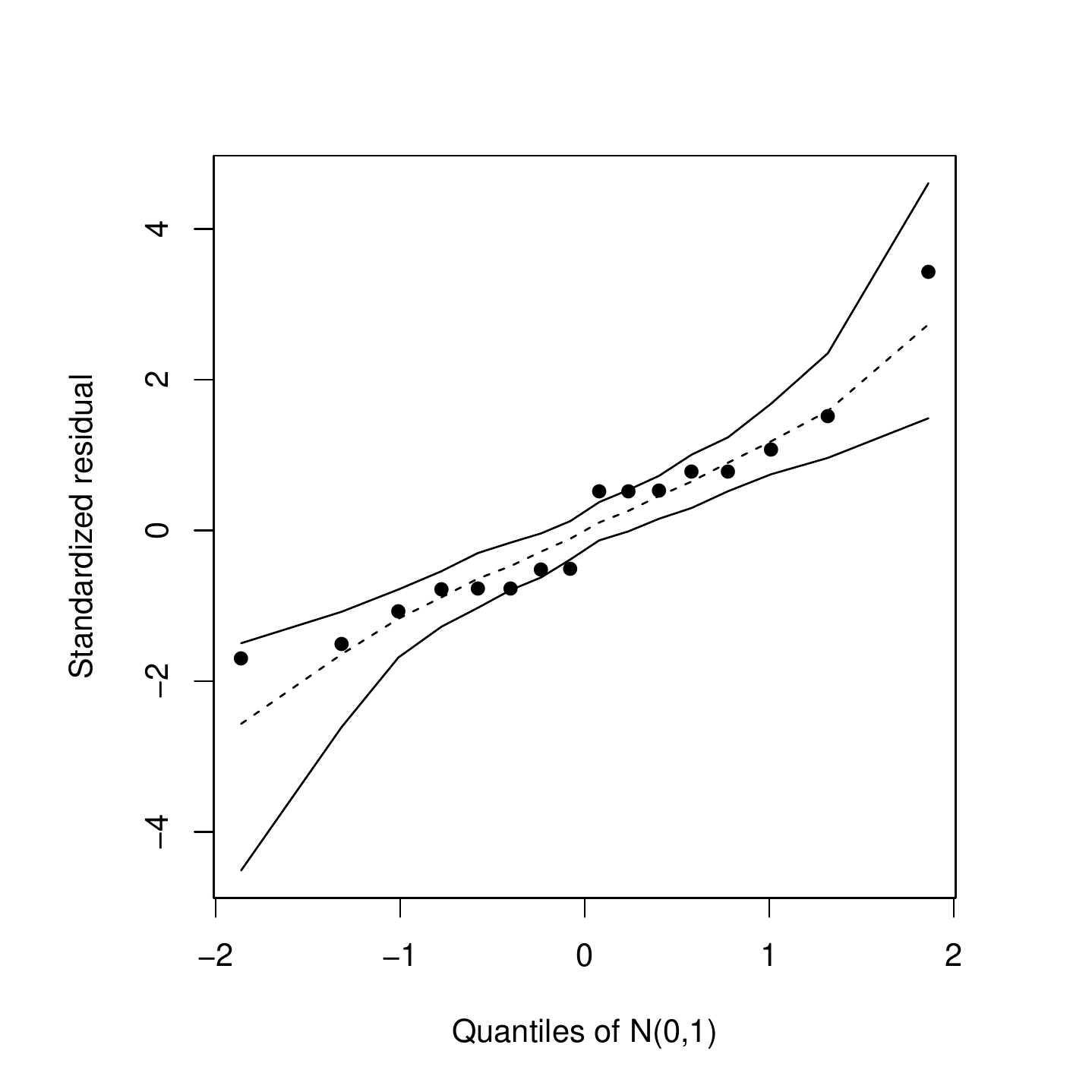}}\hspace{0.6cm}
\caption{Normal quantile-quantile plots of standardized residuals for model (\ref{modelo_queijo:1}): log-normal model (a), log-Student-t model (b), and type II log-logistic model (c); Cheddar cheese data.}
\label{fig:envelope_queijo}
\end{figure}

We now present applications of all the tests in two data sets to illustrate the need for corrections in small samples. First, we deal with the data set presented in \cite{Nateghi2012}. The aim is to investigate the effect of fat replacers on texture properties of Cheddar cheeses. Here, the response variable is the cohesiveness ($t_l$) of the cheese and the covariates are the percentage of fat ($x_{1} = 1.25\%$ -- low-fat cheese, and $2.00\%$ -- reduced-fat cheese), percentage of xanthan gum ($x_{2l} = 0.030\%$ and $0.045\%$), and percentage of sodium caseinate ($x_{3l} = 0.00\%$ and $0.15\%$). Observations were taken in $n=16$ samples of cheese in a full factorial design with two replicates; see Tables 1 and 6 in \cite{Nateghi2012}. We fit the following log-symmetric linear regression model
\begin{equation}\label{modelo_queijo:1}
\log(t_{l})=\beta_0 + \beta_1x_{1l} + \beta_2x_{2l} + \beta_3x_{3l} + \beta_4x_{1l}x_{2l} + \beta_5x_{1l}x_{3l} 
           + \beta_6x_{2l}x_{3l} + \phi\epsilon_l, 
\end{equation}
for $l=1, \ldots, 16$, where $\epsilon_l \sim S(0,1)$ are independent random errors. We considered the following standard symmetric distributions for the errors: normal, Student-t with different values for the degrees of freedom parameter, and type II logistic. The corrected AIC criteria \citep{BurnhamAnderson2004} for the fitted models are: $-135.25$ (normal), $-134.87$ (Student-t with $\nu=3$), $-135.02$ (Student-t with $\nu=4$), $-135.03$ (Student-t with $\nu=5$), and $-134.98$ (type II log-logistic). The smallest AIC is achieved by the log-normal model. Among the log-Student-t models the smallest AIC corresponds to $\nu=5$. Figure \ref{fig:envelope_queijo} shows the normal quantile-quantile plot with simulated enveloped for the standardized residuals proposed by \citet{Villegas-et-al-2013} for the log-normal model, log-Student-t model with $\nu=5$, and type II log-logistic model. Figure \ref{fig:envelope_queijo} and the AIC criteria suggest that the model that best fits the data is the log-normal linear regression model. This is the model chosen for the analysis that follow. The maximum likelihood estimates of the parameters (standard errors in parentheses) are: $\widehat{\beta}_{0} = -0.1321\,(0.0067)$, $\widehat{\beta}_{1} = -0.0043\,(0.0039)$, $\widehat{\beta}_{2} = -0.1456\,(0.1712)$,  $\widehat{\beta}_{3} = 0.0135\,(0.0251)$, $\widehat{\beta_{4}}=-0.1864\,(0.1001)$, $\widehat{\beta_{5}}=-0.0074\,(0.0100)$, $\widehat{\beta_{6}}= 0.6606\,(0.5007)$, and $\widehat{\phi}=0.0011\,(0.0002)$.

We first test each interaction effect, i.e. the null hypotheses of interest are ${\cal H}_{0}: \beta_{4}=0$, ${\cal H}_{0}: \beta_{5}=0$, and ${\cal H}_{0}: \beta_{6}=0$. The test statistics ($p$-values in parentheses) for testing ${\cal H}_{0}: \beta_{4}=0$ are: $\SW=3.4632$ (0.0628), $\SLR=3.1350$ (0.0766), $\SR=2.8470$ (0.0915), $\ST=2.8469$ (0.0915), $\SLR^*=1.6654$ (0.1969), $\SR^*=1.7657$ (0.1839), and $\ST^*=1.7657$ (0.1839). The $p$-values for all the bootstrapped tests are $0.2036$.
Note that the $p$-values vary from $6.3\%$ to $20.4\%$. Although none of the tests reject the null hypothesis at the $10\%$ nominal level, at the $5\%$ nominal level the uncorrected tests lead to rejection of ${\cal H}_{0}$ unlike the analytically corrected and bootstrapped tests. As evidenced by our simulations, the uncorrected tests tend to be liberal in small samples (recall that $n=16$), and the modified tests are less size distorted. The null hypotheses ${\cal H}_{0}: \beta_{5}=0$ and ${\cal H}_{0}: \beta_{6}=0$ are not rejected by none of the tests for all the usual significance levels (all the $p$-values are greater than 45\% and 18\% for the test of ${\cal H}_{0}: \beta_{5}=0$ and ${\cal H}_{0}: \beta_{6}=0$, respectively). We now test the hypothesis of no joint interactions effect, i.e. ${\cal H}_{0}:\beta_{4}=\beta_{5}=\beta_{6}=0$. None of the tests rejects  ${\cal H}_{0}$ at the usual significance levels (all the $p$-values are greater than 12\%).

\begin{figure}[!ht]
\centering
\includegraphics[scale=0.5]{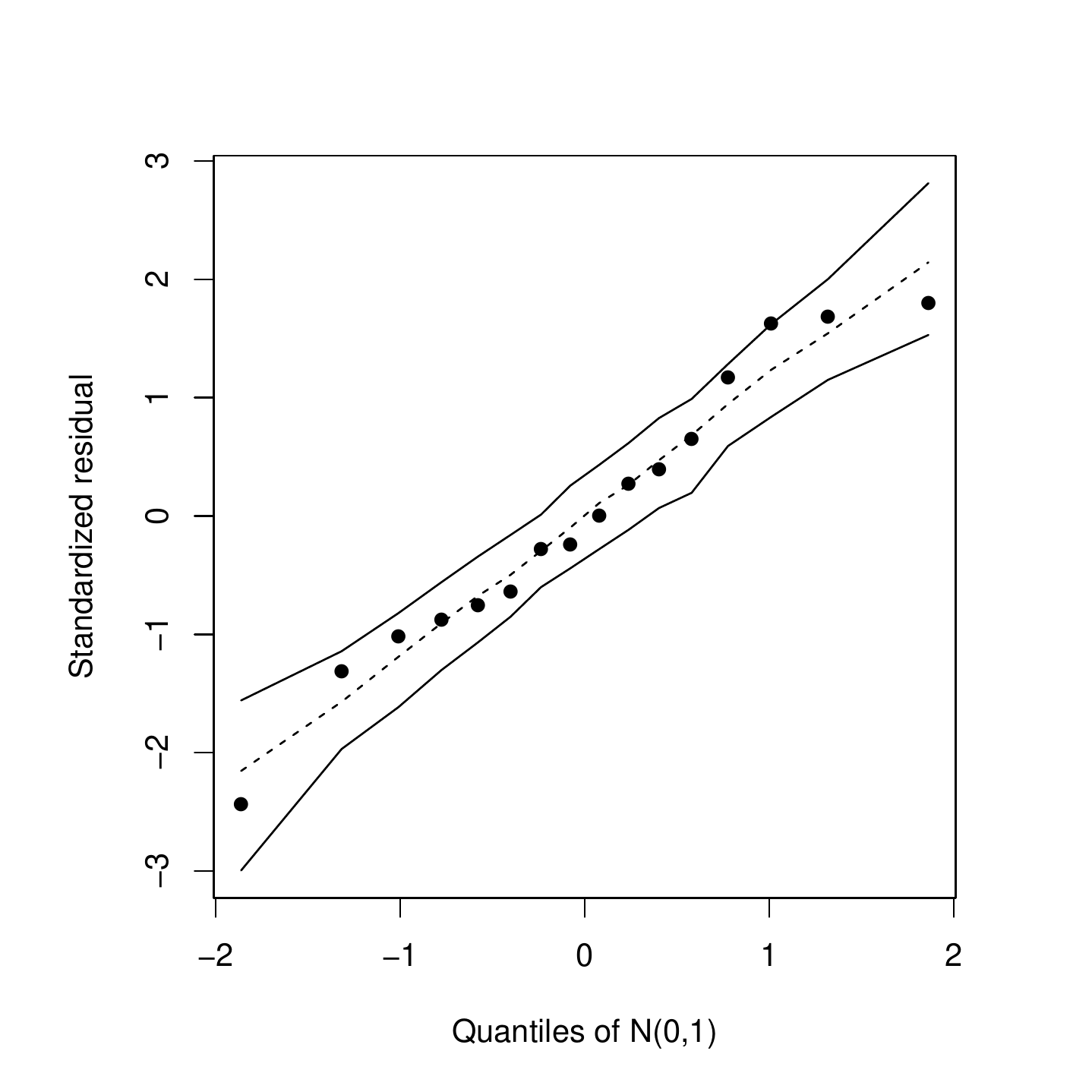}
\caption{Normal quantile-quantile plot of standardized residuals for model (\ref{modelo_queijo:final}); Cheddar cheese data.}
\label{fig:env_queijo_final}
\end{figure}

We now remove the interaction effects in model (\ref{modelo_queijo:1}) and estimate the model
\begin{align}\label{modelo_queijo:final}
\log(t_{l})&= \beta_0 + \beta_1x_{1l} + \beta_2x_{2l} + \beta_3x_{3l} + \phi\epsilon_l, \qquad l=1, \ldots, 16.
\end{align}
The maximum likelihood estimates (asymptotic standard errors in parentheses) are: $\widehat{\beta_0}=-0.1217\,$ $(0.0022)$, $\widehat{\beta_1}=-0.0119\,(0.0009)$, $\widehat{\beta_2}=-0.3989\,(0.0438)$, $\widehat{\beta_3}=0.0262\,(0.0044)$, and $\widehat{\phi}=0.0013\,$ $(0.0002)$. The null hypotheses ${\cal H}_{0}: \beta_{1}=0$, ${\cal H}_{0}: \beta_{2}=0$, and ${\cal H}_{0}: \beta_{3}=0$ are strongly rejected by all the tests at the usual significance levels. Figure \ref{fig:env_queijo_final} shows the normal quantile-quantile plot of the standardized residuals for model (\ref{modelo_queijo:final}). The plot suggests a reasonable fit. Hence, the final estimated model for the median cohesiveness of the low-fat and reduced-fat Cheddar cheese is
\begin{align*}
\widehat{\eta}= e^{-0.1217 - 0.0119x_{1} - 0.3989x_{2} + 0.0262x_{3}}.
\end{align*}

The second application considers the data set presented in Table 1 of \cite{MirhosseiniTan2010}. The data were collected to investigate the effect of emulsion components on orange beverage emulsion properties. The independent variables are the amount of gum arabic ($x_{1}$), xanthan gum ($x_{2}$) and orange oil ($x_{3}$), all measured in $g/100g$, and the response variable is the emulsion density ($y$) measured in $g/cm^{3}$. We fitted the following regression model
\begin{equation}\label{modelo_emul:inicial}
y_l=\beta_0 + \beta_1x_{1l} + \beta_2x_{2l} + \beta_3x_{3l} + \beta_4x_{1l}x_{2l} + \beta_5x_{1l}x_{3l}
   + \beta_{6}x_{2l}x_{3l} + \phi\epsilon_l, 
\end{equation}
for $l=1, \ldots, 20$, where $\epsilon_l\sim S(0,1)$ are independent errors. Different choices of the error distribution were considered as in the first application. The corrected AIC for the fitted models are: $-150.03$ (normal), $-158.68$ (Student-t with $\nu=3$), $-156.37$ (Student-t with $\nu=4$), $-154.48$ (Student-t with $\nu=5$), and $-151.03$ (type II logistic). Normal quantile-quantile plots of standardized residuals (not shown) 
%Figura \ref{fig:envelope} 
and the corrected AICs point to the Student-t model with $\nu=3$ as the best model.  The maximum likelihood estimates of the parameters (asymptotic standard errors in parentheses) are: $\widehat{\beta}_{0} = 0.9388\,(0.0220)$, $\widehat{\beta}_{1} = 0.0054\,(0.0011)$, $\widehat{\beta}_{2} = 0.1563\,(0.0401)$, $\widehat{\beta}_{3} = 0.0016\,(0.0016)$, $\widehat{\beta}_{4} = -0.0048\,(0.0015)$, $\widehat{\beta}_{5} = -0.0001\,(0.0001)$,
$\widehat{\beta}_{6} = -0.0070\,(0.0026)$, and $\widehat{\phi}=0.0012\,(0.0003)$.

%\begin{figure}[!ht]
%\centering
%\subfigure[normal\label{fig:pri_emul}]{\includegraphics[scale=0.35]{fig/envel_emul_normal_allcov_y5.pdf}}\hspace{0.6cm}
%\subfigure[t-Student\label{fig:seq_emul}]{\includegraphics[scale=0.35]{fig/envel_emul_t3_allcov_y5.pdf}}
%\subfigure[logístico II\label{fig:ter_emul}]{\includegraphics[scale=0.35]{fig/envel_emul_logistica_allcov_y5.pdf}}\hspace{0.6cm}
%\caption{Gráfico normal de probabilidades e envelope simulado para as distribuições (a) normal, (b) t-Student e (c) logística II ajustadas ao modelo (\ref{modelo_emul:inicial}).}
%\label{fig:envelope}
%\end{figure}

We first test the individual interaction effects, i.e. the null hypotheses under test are ${\cal H}_{0}: \beta_{4}=0$, ${\cal H}_{0}: \beta_{5}=0$, and ${\cal H}_{0}: \beta_{6}=0$; see Table \ref{tab_emul:beta_7_8_9} for the test statistics and $p$-values. The null hypothesis ${\cal H}_{0}: \beta_{4}=0$ is rejected by the Wald, likelihood ratio and gradient tests at the $5\%$ nominal level. However, the opposite decision is reached by the score and the modified tests. None of the tests reject ${\cal H}_{0}: \beta_{5}=0$, and ${\cal H}_{0}: \beta_{6}=0$ is rejected by the Wald and likelihood ratio tests at the $5\%$ nominal level, but is not rejected by the others. It is noticeable that there is no conflict among the modified tests, and all of them do not show enough evidence to reject the null hypotheses. We then test the joint interactions effect in model (\ref{modelo_emul:inicial}), i.e. the null hypothesis is   ${\cal H}_{0}: \beta_{4}=\beta_{5}=\beta_{6}=0$. The test statistics ($p$-values in parentheses) are: $\SW=17.9297$ (0.0005), $\SLR=8.1531$ (0.0430), $\SR=2.9646$ (0.3971), $\ST= 4.5884$ (0.2045), $\SLR^*=4.0333$ (0.2579), $\SR^*=2.1259$ (0.5467), and  $\ST^*=3.1251$ (0.3727). The $p$-values of the bootstrapped Wald, likelihood ratio, score and gradient tests are $0.1838$, $0.2280$, $0.5868$, and $0.3466$, respectively.
Note that the $p$-values range from $0.05\%$ (Wald) to  $58.7\%$ (bootstrapped score test). While the Wald and the likelihood ratio tests reject the joint interactions effect, the other tests point to the opposite direction. 

%\begin{table}[!htp]
%\begin{center}
%\caption{Test statistics ($p$-values in parentheses) for testing ${\cal H}_{0}: \beta_{4}=0$, ${\cal H}_{0}: \beta_{5}=0$, and ${\cal H}_{0}: \beta_{6}=0$ in model (\ref{modelo_emul:inicial}); orange emulsion data.}\label{tab_emul:beta_7_8_9}
%\begin{footnotesize}
%\begin{tabular}{lccccccccccc}
%\hline
%  null hypothesis & $\SW$ &  $\SLR$ & $\SR$ &  $\ST$ & $\SLR^*$ & $\SR^*$ & $\ST^*$ & $\SW^b$ &  $\SLR^b$ & $\SR^b$ &  $\ST^b$\\ 
%\hline        												
%    ${\cal H}_{0}:\beta_{4}=0$  & 10.2240 & 6.5050 & 3.5812 & 4.1713 & 2.5065 & 2.2753 & 2.1510 & & & & 
%                                \\  
%		                            & (0.0014) & (0.0108) & (0.0584) & (0.0411) & (0.1134) & (0.1314) & (0.1425)
%																& (0.1050) & (0.0922) & (0.1548) & (0.1306)
%																\\
%	  ${\cal H}_{0}:\beta_{5}=0$  & 0.4583 & 0.5354 & 0.6040 & 0.5148 &  0.2063 & 0.3500 & 0.2066 & & & &
%		                            \\  
%		                            & (0.4984) & (0.4644) & (0.4371) & (0.4731) & (0.6497) & (0.5541) & (0.6494)                                & (0.6728) & (0.6246) & (0.5510) & (0.6196)
%																\\
%		${\cal H}_{0}:\beta_{6}=0$  & 7.2474 & 5.2526 & 3.3333 & 3.5959 & 2.0239 & 2.1023 & 1.7897 & & & &
%		                            \\  
%		                            & (0.0071) & (0.0219) & (0.0679) & (0.0579) & (0.1548) & (0.1471) & (0.1810)                                & (0.1532) & (0.1266) & (0.1690) & (0.1650)

%																\\		
%\hline
%\end{tabular}
%\end{footnotesize}
%\end{center}
%\end{table}

\begin{table}[!htp]
\centering
{\footnotesize
\caption{Test statistics and $p$-values  for testing ${\cal H}_{0}: \beta_{4}=0$, ${\cal H}_{0}: \beta_{5}=0$, and ${\cal H}_{0}: \beta_{6}=0$ in model (\ref{modelo_emul:inicial}); orange emulsion data.}\label{tab_emul:beta_7_8_9}
\begin{tabular}{lcccccc}  \hline
                        & \multicolumn{2}{c}{${\cal H}_{0}: \beta_{4}=0$} & \multicolumn{2}{c}{${\cal H}_{0}: \beta_{5}=0$}  & \multicolumn{2}{c}{${\cal H}_{0}: \beta_{6}=0$}\\
 statistic             &  observed value   & $p-$value  &  observed value   & $p-$value &  observed value   & $p-$value \\   \hline
  ${S}_\W$             & 10.2240& 0.0014 & 0.4583 & 0.4984 & 7.2474  & 0.0071 \\
  ${S}_{\LR}$          & 6.5050 & 0.0108 & 0.5354 & 0.4644 & 5.2526 & 0.0219 \\
  ${S}_{\R}$           & 3.5812 & 0.0584 & 0.6040 & 0.4371 & 3.3333 & 0.0679\\
  ${S}_{\T}$           & 4.1713 & 0.0411 & 0.5148 & 0.4731 & 3.5959 & 0.0579 \\
 ${S}^{*}_{\LR}$       & 2.5065 & 0.1134 & 0.2063 & 0.6497 & 2.0239 & 0.1548\\
  ${S}^{*}_{\R}$       & 2.2753 & 0.1314 & 0.3500 & 0.5541 & 2.1023 & 0.1471 \\
  ${S}^{*}_{\T}$       & 2.1510 & 0.1425 & 0.2066 & 0.6494 & 1.7897 & 0.1810\\
  ${S}^{b}_{\W}$       &        & 0.1054 &        & 0.6728 &        & 0.1532\\
  ${S}^{b}_{\LR}$      &        & 0.0930 &        & 0.6242 &        & 0.1266 \\
  ${S}^{b}_{\R}$       &        & 0.1562 &        & 0.5506 &        & 0.1690  \\
  ${S}^{b}_{\T}$       &        & 0.1318 &        & 0.6194 &        &  0.1650 \\  \hline
\end{tabular}
}      
\end{table}

Removing the interaction effects in (\ref{modelo_emul:inicial}) we now estimate the model
\begin{align}\label{mr:emulsao2}
\begin{split}
y_l&=\beta_0 + \beta_1x_{1l} + \beta_2x_{2l} + \beta_3x_{3l} + \phi\epsilon_l.
\end{split}
\end{align}
The maximum likelihood estimates  (asymptotic standard errors in parentheses) are $\widehat{\beta_0}= 1.0198\,$ $(0.0051)$, $\widehat{\beta_1}= 0.0027\,(0.0002)$, $\widehat{\beta_2}=-0.0058\,(0.0059)$, $\widehat{\beta_3}=-0.0023\,(0.0003)$, and $\widehat{\phi}=0.0018\,$ $(0.0004)$. At the usual nominal significance levels, all the tests strongly reject ${\cal H}_{0}: \beta_{1}=0$ and ${\cal H}_{0}: \beta_{3}=0$. Also, all the tests suggest the removal of $x_2$ from model (\ref{mr:emulsao2}). Hence, the final model is
%\begin{align}\label{modelo_emul:final}
%\begin{split}
$
y_l=\beta_0 + \beta_1x_{1l} + \beta_3x_{3l} + \phi\epsilon_l.
$
%\end{split}
%\end{align}
The maximum likelihood estimates of the parameters are $\widehat{\beta}_{0} = 1.0168\,(0.0047)$, $\widehat{\beta}_{1} =  0.0027\,(0.0002)$, $\widehat{\beta}_{3} = -0.0023\,(0.0003)$, and $\widehat{\phi}=  0.0018\,(0.0004)$. The normal quantile-quantile plot of standardized residuals for the final estimated model (not shown) suggests a suitable fit.

%Figure \ref{fig:env_emul_final_1} shows the normal probability plot of standardized residuals for model (\ref{modelo_emul:final}). 

%\begin{figure}[!ht]
%\centering
%\includegraphics[scale=0.5]{fig/envel_emul_t3_final_y5.pdf}
%\caption{Gráfico normal de probabilidades e envelope simulado para o modelo (\ref{modelo_emul:final}).}
%\label{fig:env_emul_final_1}
%\end{figure}

%%%%%%%%%%%%%%%%%%%%%%%%%%%%%%%%%%%%%%%%%%%%%%%%%%%%%%%%%%%%%%%%%%%%%%%%%%%%%%%%%%%%%%%%%%%%%%%%%%%%%%%%%%%%%%%%%%%%%%%%%%%%%%%%%%%%%%%%%%%%
%                                            Final remarks
%%%%%%%%%%%%%%%%%%%%%%%%%%%%%%%%%%%%%%%%%%%%%%%%%%%%%%%%%%%%%%%%%%%%%%%%%%%%%%%%%%%%%%%%%%%%%%%%%%%%%%%%%%%%%%%%%%%%%%%%%%%%%%%%%%%%%%%%%%%%%

\section{Final remarks}\label{MRLS-sec:7} 

This paper dealt with the issue of testing hypotheses in symmetric and log-symmetric linear regression models. The models can be easily fitted using the available package {\tt ssym} in {\tt R}. Testing inference using the classic tests and the recently proposed gradient test rely on asymptotic approximations and may be unreliable when the sample size is small or even moderate. Our simulations indicate that the Wald and the likelihood tests may be severely liberal in finite samples. The score and the gradient tests are less size distorted but may present considerable size distortion depending on the number of observations, regression parameters, and parameters under test. 

We derived a Bartlett-type correction to the gradient statistic in symmetric linear regression models. We showed that this correction and the corrections to the likelihood ratio and score statistics found in the literature are also valid for log-symmetric linear regression models. We then performed simulation experiments comparing the uncorrected tests and their corresponding analytically corrected (except for the Wald test) and bootstrapped versions. The simulations are clear in indicating that the modified tests are much less size distorted than the original tests and that all the tests have similar power. The analytical corrections are simple and easily implemented is any software that performs matrix computation, such as {\tt R}. The bootstrapped tests, on the other hand, requires computationally-intensive calculations. Since there is no Bartlett-type correction available to the Wald statistic, the bootstrap method is convenient when performing Wald tests. 

We presented two applications for real data. Our analyses illustrate that the use of the original tests may be misleading in small samples. The usefulness of the analytically corrected and bootstrapped tests became clear. We, therefore, recommend the use of the modified tests when performing testing inference in symmetric and log-symmetric linear regression models.

%%%%%%%%%%%%%%%%%%%%%%%%%%%%%%%%%%%%%%%%%%%%%%%%%%%%%%%%%%%%%%%%%%%%%%%%%%%%%%%%%%%%%%%%%%%%%%%%%%%%%%%%%%%%%%%%%%%%%%%%%%%%%%%%%%%%%%%%%%%%
%                                           Acknowledgments
%%%%%%%%%%%%%%%%%%%%%%%%%%%%%%%%%%%%%%%%%%%%%%%%%%%%%%%%%%%%%%%%%%%%%%%%%%%%%%%%%%%%%%%%%%%%%%%%%%%%%%%%%%%%%%%%%%%%%%%%%%%%%%%%%%%%%%%%%%%%%

\vspace{0.5cm}
\section*{Acknowledgments}

We gratefully acknowledge grants from the Brazilian agencies CNPq and FAPESP.

%%%%%%%%%%%%%%%%%%%%%%%%%%%%%%%%%%%%%%%%%%%%%%%%%%%%%%%%%%%%%%%%%%%%%%%%%%%%%%%%%%%%%%%%%%%%%%%%%%%%%%%%%%%%%%%%%%%%%%%%%%%%%%%%%%%%%%%%%%%%
%                                          Appendix
%%%%%%%%%%%%%%%%%%%%%%%%%%%%%%%%%%%%%%%%%%%%%%%%%%%%%%%%%%%%%%%%%%%%%%%%%%%%%%%%%%%%%%%%%%%%%%%%%%%%%%%%%%%%%%%%%%%%%%%%%%%%%%%%%%%%%%%%%%%%%

\section*{Appendix}

Let $\kappa_{rs}=\Es(\partial^{2} \ell/\partial\beta_r\partial\beta_s)$, $\kappa_{rst}=\Es(\partial^{3} \ell/\partial\beta_r\partial\beta_s\partial\beta_t)$, $\kappa_{rstu}=\Es(\partial^{4} \ell/\partial\beta_r\partial\beta_s\partial\beta_t\partial\beta_u)$, $\kappa_{rs}^{(t)}=\partial \kappa_{rs}/\partial \beta_{t}$, $\kappa_{rs}^{(tu)}=\partial^{2} \kappa_{rs}/\partial\beta_t\partial\beta_u$, $\kappa_{\phi\phi}=\Es(\partial^{2} \ell/\partial\phi^2)$, $\kappa_{r\phi}=\Es(\partial^{2} \ell/\partial\beta_r\partial\phi)$, $\kappa_{r\phi}^{(s)}=\partial^{2} \kappa_{r\phi}/\partial\beta_s$ and so on. The indices $r$, $s$, $t$ and $u$ vary from $1$ to $p$. 
In symmetric linear regression models we have
\begin{align*}\label{kappas}
%\begin{split}
\kappa_{rs}&=\frac{\delta_{01000}}{\phi^2}\sum_{l=1}^{n}x_{lr}x_{ls}, \qquad \kappa_{rstu}=\frac{\delta_{00010}}{\phi^4}\sum_{l=1}^{n}x_{lr}x_{ls}x_{lu}x_{lt},\qquad
\kappa_{rst}=\kappa_{rs}^{(t)}=\kappa_{rst}^{(u)}=\kappa_{rs}^{(tu)}=0,\\
\kappa_{\phi\phi}&= -\frac{n}{\phi^2}(1-\delta_{20002}), \qquad
\kappa^{(\phi)}_{\phi\phi}= -\frac{2n}{\phi^3}(\delta_{01002}-1),\qquad \kappa_{\phi\phi\phi} = -\frac{n}{\phi^3}(6\delta_{01002}+\delta_{00103}-4),\\
\kappa_{r\phi}&=\kappa_{r\phi}^{(s)}=\kappa_{r\phi\phi}=\kappa^{(u)}_{r\phi\phi}=0,\qquad
\kappa_{rs\phi}= -\frac{1}{\phi^3}(\delta_{00101}+2\delta_{01000})\sum_{l=1}^{n} x_{lr}x_{ls}, \\
\kappa_{rs\phi\phi}&= -\frac{1}{\phi^4}(\delta_{00012}-6\delta_{11001})\sum_{l=1}^{n} x_{lr}x_{ls},\qquad
\kappa^{(\phi)}_{rs\phi}= \frac{3}{\phi^4}(\delta_{00101}+2\delta_{01000})\sum_{l=1}^{n} x_{lr}x_{ls}, \\
\kappa^{(\phi)}_{rs}&= -\frac{2\delta_{01000}}{\phi^3}\sum_{l=1}^{n} x_{lr}x_{ls}, 
%\end{split}
\end{align*}
see \cite{Uribe2001} and \cite{Uribe2007}.

Let $\bm{K}^{-1}=\diag\{\bm{K}^{-1}_{\bm\beta}, -\kappa^{-1}_{\phi\phi}\}$ be the Fisher information matrix inverse of $(\bm{\beta}_{1}^{\top},\bm{\beta}_{2}^{\top}, \phi)^{\top}$. Let
\[
{\bm{\mathcal{A}}}_{\bm{\beta}} =
\begin{bmatrix}
\bm{0} & \bm{0} \\
\bm{0} & \bm{K}_{\bm{\beta}_{22}}^{-1}
\end{bmatrix},
\qquad
{\bm{\mathcal{M}}}_{\bm{\beta}} = \bm{K}^{-1}_{\bm{\beta}} - {\bm{\mathcal{A}}}_{\bm\beta},
\] 
${\bm{\mathcal{A}}}_{\vecbeta\phi}=\diag\{{\bm{\mathcal{A}}}_{\bm{\beta}}, -\kappa^{-1}_{\phi\phi}\}$, and ${\bm{\mathcal{M}}}_{\vecbeta\phi}=\diag\{{\bm{\mathcal{M}}}_{\bm{\beta}},0\}$. We denote by $m^{r\phi}$ and $a^{r\phi}$ the element $(r,p+1)$ of $\bm{\mathcal{M}}_{\bm{\beta}\phi}$ and $\bm{\mathcal{A}}_{\bm{\beta}\phi}$, respectively. Analogously, $m^{\phi\phi}$ and $a^{\phi\phi}$ represent the element $(p+1,p+1)$ of $\bm{\mathcal{M}}_{\bm{\beta}\phi}$ and $\bm{\mathcal{A}}_{\bm{\beta}\phi}$, respectively. We have $m^{r\phi}=m^{\phi r}=m^{\phi\phi}=a^{r\phi}=a^{\phi r}=0$, for $r=1,\ldots,p$, and $a^{\phi\phi}=-\kappa_{\phi\phi}^{-1}$.

The coefficients $A_{\T}$'s that define the Bartlett-type correction to the gradient statistic in symmetric linear regression models are obtained by replacing the moments above in the formulas for the $A$'s in Theorem 1 of \cite{VargasFerrariLemonte2013}. We first note that $A_{\T11}=A_{\T1}+A_{\T1,\bm{\beta}\phi}$, $A_{\T22}=A_{\T2}+A_{\T2,\bm{\beta}\phi}$, and $A_{\T33}=A_{\T3}+A_{\T3,\bm{\beta}\phi}$, where $A_{\T1}$, $A_{\T2}$, and $A_{\T3}$ are the coefficients obtained assuming that $\phi$ is known and $A_{\T1,\bm{\beta}\phi}$, $A_{\T2,\bm{\beta}\phi}$, and $A_{\T3,\bm{\beta}\phi}$ are the additional terms that appear when $\phi$ is unknown.

By replacing the moments above in the formula of $A_1$ in Theorem 1 of \cite{VargasFerrariLemonte2013}, the coefficient $A_{\T1}$ can be written as
\begin{align*}
{A}_{\T1}=6\sum\nolimits^{\prime}\kappa_{jrsu}m^{jr}a^{su}=\frac{6\delta_{00010}}{\phi^4}\sum\nolimits^{\prime}\sum_{l=1}^{n}x_{lr}x_{lj}x_{ls}x_{lu}m^{jr}a^{su},
\end{align*}
where $\sum\nolimits^{\prime}$ is the summation over the indices of the parameter $\vecbeta$.  
Inverting the order of the summations and rearranging the terms we have
\begin{align*}
{A}_{\T1}=\frac{6\delta_{00010}}{\phi^4}\sum_{l=1}^{n}\left(\sum\nolimits^{\prime}x_{lj}m^{jr}x_{lr}\right)\left(\sum\nolimits^{\prime}x_{ls}a^{su}x_{lu}\right).
\end{align*}
The terms $\sum\nolimits^{\prime}x_{li}a^{ij}x_{lj}$ and $\sum\nolimits^{\prime}x_{li}m^{ij}x_{lj}$ represent the $(l,l)$ element of the matrices $\phi^2\bm{Z}_2/\delta_{20000}$ and $\phi^2(\bm{Z}-\bm{Z}_2)/\delta_{20000}$, respectively. Hence,
\[
{A}_{\T1}=%\frac{6\delta_{00010}}{\phi^4}\sum_{l=1}^{n}\left(\frac{\phi^2}{\delta_{20000}}(z_{ll}-z_{2ll})\right)\left(\frac{\phi^2}{\delta_{20000}}z_{2ll}\right)
=\frac{6\delta_{00010}}{\delta^2_{20000}}\sum_{l=1}^{n}(z_{ll}-z_{2ll})z_{2ll}%=6c_0\bm{1}_{n}^{\top}\bm{Z}_{2d}(\bm{Z}_d-\bm{Z}_{2d})\bm{1}_{n},
=\frac{6c_0}{n} (\rho_{ZZ_2}-\rho_{Z_2Z_2}),
\]
where $c_0$ and the $\rho$'s are given in Section \ref{MRLS-sec:4}.

Analogously, we have
\begin{align*}
{A}_{\T2}&=-3\sum\nolimits^{\prime}\kappa_{jrsu}m^{jr}m^{su}=-\frac{3\delta_{00010}}{\phi^4}\sum_{l=1}^{n}\left(\sum\nolimits^{\prime}x_{lj}m^{jr}x_{lr}\right)\left(\sum\nolimits^{\prime}x_{ls}m^{su}x_{lu}\right)\\
%&=-\frac{3\delta_{00010}}{\phi^4}\sum_{l=1}^{n}\left(\frac{\phi^2}{\delta_{20000}}(z_{ll}-z_{2ll})\right)\left(\frac{\phi^2}{\delta_{20000}}(z_{ll}-z_{2ll})\right)\\
&=-\frac{3\delta_{00010}}{\delta^2_{20000}}\sum_{l=1}^{n}(z_{ll}-z_{2ll})^2=-\frac{3c_0}{n}(\rho_{ZZ}-2\rho_{ZZ_2}+\rho_{Z_2Z_2})
%-3c_0\bm{1}_{n}^{\top}(\bm{Z}_d-\bm{Z}_{2d})^2\bm{1}_{n}
\end{align*}
and ${A}_{\T3}=0.$

We now turn to the derivation of $A_{\T1,\bm{\beta}\phi}$, $A_{\T2,\bm{\beta}\phi}$, and $A_{\T3,\bm{\beta}\phi}$. These terms only appear when $\phi$ is unknown. When $\phi$ is unknown, which is usually the case, it follows from Theorem 1 of \cite{VargasFerrariLemonte2013} that $A_{\T3,\bm{\beta}\phi}=0$, 
\begin{align}  \label{parc1}
\begin{split}
{A}_{\T1,\bm{\beta}\phi}&=\sum\nolimits^{\prime}\kappa_{jrs}\kappa_{u\phi\phi}a^{\phi\phi}m^{jr}m^{su}+ \sum\nolimits^{\prime}\kappa_{jrs}\kappa_{u\phi\phi}a^{\phi\phi}m^{js}m^{ru} \\
&\quad+\sum\nolimits^{\prime}\kappa_{jrs}\kappa_{u\phi\phi}a^{\phi\phi}m^{ju}m^{rs}
+6\sum\nolimits^{\prime}\kappa_{jr\phi}\kappa_{\phi\phi\phi}m^{jr}(a^{\phi\phi})^2 \\
&\quad+6\sum\nolimits^{\prime}\kappa_{jr\phi}\kappa_{vw\phi}m^{jr}a^{vw}a^{\phi\phi}
+6\sum\nolimits^{\prime}\kappa_{jrs}\kappa_{u\phi\phi}m^{jr}a^{su}a^{\phi\phi} \\
&\quad+3\sum\nolimits^{\prime}\kappa_{j\phi\phi}\kappa_{uvw}m^{ju}a^{vw}a^{\phi\phi}+3\sum\nolimits^{\prime}\kappa_{jrs}\kappa_{u\phi\phi}m^{ju}a^{rs}a^{\phi\phi} \\
&\quad+3\sum\nolimits^{\prime}\kappa_{j\phi\phi}\kappa_{u\phi\phi}m^{ju}(a^{\phi\phi})^{2}
+ 3\sum\nolimits^{\prime}\kappa_{jr\phi}\kappa_{uw\phi}m^{ju}a^{rw}a^{\phi\phi} \\
&\quad+3\sum\nolimits^{\prime}\kappa_{js\phi}\kappa_{uv\phi}m^{ju}a^{sv}a^{\phi\phi}+3\sum\nolimits^{\prime}\kappa_{jr\phi}\kappa_{uv\phi}m^{ju}a^{rv}a^{\phi\phi} \\
&\quad+3\sum\nolimits^{\prime}\kappa_{js\phi}\kappa_{uw\phi}m^{ju}a^{sw}a^{\phi\phi}
- 6\sum\nolimits^{\prime}(\kappa_{jr\phi}^{(\phi)}-\kappa_{jr\phi\phi})m^{jr}a^{\phi\phi} \\
&\quad- 6\sum\nolimits^{\prime}\kappa_{jr\phi}^{(\phi)}m^{jr}a^{\phi\phi}
 -12\sum\nolimits^{\prime}\kappa_{j\phi\phi}^{(u)}m^{ju}a^{\phi\phi} \\ &\quad-12\sum\nolimits^{\prime}\kappa_{kl}^{(\phi)}\kappa_{jr\phi}(\kappa^{jk}\kappa^{lr}-a^{jk}a^{lr})a^{\phi\phi}
-12\sum\nolimits^{\prime}\kappa_{\phi\phi}^{(\phi)}\kappa_{jr\phi}m^{jr}(a^{\phi\phi})^{2} \\
&\quad-12\sum\nolimits^{\prime}\kappa_{j\phi\phi}\kappa_{kl}^{(u)}(\kappa^{jk}\kappa^{lu}-a^{jk}a^{lu})a^{\phi\phi},
\end{split}
\end{align}
and
\begin{align}  \label{parc2}
\begin{split}
{A}_{\T2,\bm{\beta}\phi}&=-\sum\nolimits^{\prime}\kappa_{jrs}\kappa_{u\phi\phi}m^{jr}m^{su}a^{\phi\phi}-\sum\nolimits^{\prime}\kappa_{jrs}\kappa_{u\phi\phi}m^{js}m^{ru}a^{\phi\phi} \\
&\quad-\sum\nolimits^{\prime}\kappa_{jrs}\kappa_{u\phi\phi}m^{ju}m^{rs}a^{\phi\phi}
-3\sum\nolimits^{\prime}\kappa_{jr\phi}\kappa_{uv\phi}m^{jr}m^{uv}a^{\phi\phi} \\
&\quad-3\sum\nolimits^{\prime}\kappa_{jr\phi}\kappa_{uv\phi}m^{ju}m^{rv}a^{\phi\phi}-3\sum\nolimits^{\prime}\kappa_{jr\phi}\kappa_{uv\phi}m^{jv}m^{ru}a^{\phi\phi}.
\end{split}
\end{align}
Plugging the $\kappa$'s in \eqref{parc1} and \eqref{parc2} some terms vanish. By inverting the summation order and rearranging the terms we have
\begin{align*} 
\begin{split}
{A}_{\T1,\bm{\beta}\phi}&=\frac{6(\delta_{00101}+2\delta_{01000})(6\delta_{01002}+\delta_{00103}-4)}{n\phi^2(\delta_{20002}-1)^2}\sum_{l=1}^{n}\left(\sum\nolimits^{\prime}x_{lj}m^{jr}x_{lr}\right)\\
&+ \frac{6(\delta_{00101}+2\delta_{01000})^2}{n\phi^4(\delta_{20002}-1)}\sum_{l,i=1}^{n}\left(\sum\nolimits^{\prime}x_{lj}m^{jr}x_{lr}\right)\left(\sum\nolimits^{\prime}x_{iv}a^{vw}x_{iw}\right)\\
&-\frac{36(\delta_{00101}+2\delta_{01000})}{n\phi^2(\delta_{20002}-1)}\sum_{l=1}^{n}\left(\sum\nolimits^{\prime}x_{lj}m^{jr}x_{lr}\right) + \frac{6(\delta_{00012}-6\delta_{11001})}{n\phi^2(\delta_{20002}-1)}\sum_{l=1}^{n}\left(\sum\nolimits^{\prime}x_{lj}m^{jr}x_{lr}\right)\\
& - \frac{24\delta_{01000}(\delta_{00101}+2\delta_{01000})}{n\phi^4(\delta_{20002}-1)}\sum_{l=1}^{n}\left(-\sum\nolimits^{\prime}x_{lj}k^{jk}x_{ik}\right)\left(-\sum\nolimits^{\prime}x_{is}k^{sr}x_{lr}\right)\\
&+ \frac{24\delta_{01000}(\delta_{00101}+2\delta_{01000})}{n\phi^4(\delta_{20002}-1)}\sum_{l=1}^{n}\left(\sum\nolimits^{\prime}x_{lj}a^{jk}x_{ik}\right)\left(\sum\nolimits^{\prime}x_{is}a^{sr}x_{lr}\right)\\
&- \frac{24(\delta_{01002}-1)(\delta_{00101}+2\delta_{01000})}{n\phi^2(\delta_{20002}-1)^2}\sum_{l=1}^{n}\left(\sum\nolimits^{\prime}x_{lj}m^{jr}x_{lr}\right)
\end{split}
\end{align*}
and 
\begin{align*} 
\begin{split}
{A}_{\T2,\bm{\beta}\phi}&=-\frac{3(\delta_{00101}+2\delta_{01000})^2}{n\phi^4(\delta_{20002}-1)}\sum_{l,i=1}^{n}\left(\sum\nolimits^{\prime}x_{lj}m^{jr}x_{lr}\right)\left(\sum\nolimits^{\prime}x_{iu}m^{uv}x_{iv}\right)\\
&- \frac{6(\delta_{00101}+2\delta_{01000})^2}{n\phi^4(\delta_{20002}-1)}\sum_{l,i=1}^{n}\left(\sum\nolimits^{\prime}x_{lj}m^{ju}x_{iu}\right)\left(\sum\nolimits^{\prime}x_{lr}a^{rv}x_{iv}\right).\\
\end{split}
\end{align*}
Note that $\sum\nolimits' x_{li}\kappa^{ij}x_{lj}$ equals the $(l,l)$ element of the matrix $\phi^2\bm{Z}/\delta_{20000}$ and that
\[
\qquad \sum\nolimits^{\prime} z_{ll} = \mathrm{tr}(Z_d)=p, \qquad \sum\nolimits^{\prime} z_{2ll}=\mathrm{tr}(Z_{2d})=p-q, \qquad \sum\nolimits^{\prime} (z_{ll}-z_{2ll}) = \mathrm{tr}(Z_d-Z_{2d})=q.
\]
Hence,
\begin{align*} 
\begin{split}
{A}_{\T1,\bm{\beta}\phi}&=\frac{6(\delta_{00101}+2\delta_{01000})(6\delta_{01002}+\delta_{00103}-4)}{n\delta_{20000}(\delta_{20002}-1)^2}q + \frac{6(\delta_{00101}+2\delta_{01000})^2}{n\delta^2_{20000}(\delta_{20002}-1)}q(p-q)\\
\\
&-\frac{36(\delta_{00101}+2\delta_{01000})}{\delta_{20000}(\delta_{20002}-1)}q + \frac{6(\delta_{00012}-6\delta_{11001})}{n\delta_{20000}(\delta_{20002}-1)}q - \frac{24\delta_{01000}(\delta_{00101}+2\delta_{01000})}{n\delta^2_{20000}(\delta_{20002}-1)}p\\
\\
&+ \frac{24\delta_{01000}(\delta_{00101}+2\delta_{01000})}{n\delta^2_{20000}(\delta_{20002}-1)}(p-q)- \frac{24(\delta_{01002}-1)(\delta_{00101}+2\delta_{01000})}{n\delta^2_{20000}(\delta_{20002}-1)^2}q
\end{split}
\end{align*}
and
\begin{align*} 
\begin{split}
{A}_{\T2,\bm{\beta}\phi}&=-\frac{3(\delta_{00101}+2\delta_{01000})^2}{n\delta^2_{20000}(\delta_{20002}-1)}\left(q^2+2q\right).
\end{split}
\end{align*}
After some algebra, we arrive at the expressions for ${A}_{\T1,\bm{\beta}\phi}$ and ${A}_{\T2,\bm{\beta}\phi}$ given in Section \ref{MRLS-sec:4}.

\end{document}